\documentclass[prd,amsmath,aps,floats,amssymb,floatfix,superscriptaddress,nofootinbib,twocolumn,preprintnumbers]{revtex4-2}
\UseRawInputEncoding
\usepackage{tabularx}
\usepackage{amsmath,amssymb,natbib,latexsym,times}
\defcitealias{PratBaxter2019}{Prat, Baxter et al.}

\usepackage[T1]{fontenc}
\usepackage{aecompl}

\usepackage{multirow}
\usepackage{rotating}
\usepackage{booktabs}
\usepackage{makecell}
\usepackage{soul}
\usepackage{enumitem}
\usepackage[utf8]{inputenc}

\newcommand{\aap}{A\&A}
\newcommand{\apjs}{ApJS}
\newcommand{\aj}{A.J.}
\newcommand{\mnras}{MNRAS}

\newcommand{\jcap}{JCAP}
\newcommand{\apjl}{ApJ}

\usepackage{verbatim}
\usepackage[usenames,dvipsnames]{xcolor}
\usepackage{tabulary}
\usepackage{tabularx}
\usepackage{todonotes}
\usepackage{subcaption}
\usepackage{hyperref}
\hypersetup{colorlinks=true, linkcolor=blue, citecolor=blue, filecolor=blue, urlcolor=blue}
\usepackage[normalem]{ulem} 

\newcommand\fig[1]{Figure~\ref{#1}}

\usepackage[dvipsnames]{xcolor}

\newcommand\be{\begin{equation}}
\newcommand\ee{\end{equation}}
\def\bea{\begin{eqnarray}}
\def\eea{\end{eqnarray}}

\newcommand\red[1]{\textcolor{red}{#1}}

\allowdisplaybreaks

\begin{document}

\title{Lensing ratios as a geometric probe of dark energy in the LSST and Simons Observatory era}


\author{Judit Prat}
\affiliation{Departament de F\'{\i}sica (FIS), Grup d'Astronomia i Astrof\'{\i}sica (GAA), Escola d'Enginyeria de Barcelona Est (EEBE), Universitat Polit\`{e}cnica de Catalunya - BarcelonaTech (UPC), Campus Diagonal-Bes\`{o}s, Av.~d'Eduard Maristany, 16, 08019 Barcelona, Spain}
\affiliation{DARK, Niels Bohr Institute, University of Copenhagen, Blegdamsvej 17, 2100 Copenhagen, Denmark}

\author{Carles S\'anchez}
\affiliation{Departament de F\'{\i}sica, Universitat Aut\`{o}noma de Barcelona (UAB), 08193 Bellaterra (Barcelona), Spain}
\affiliation{Institut de F\'{\i}sica d'Altes Energies (IFAE), The Barcelona Institute of Science and Technology, Campus UAB, 08193 Bellaterra (Barcelona), Spain}

\date{\today}
\label{firstpage}

\begin{abstract}

Ratios of galaxy--CMB and galaxy--galaxy lensing around the same foreground galaxies probe relative distances extending to recombination without a standard candle or ruler. For sufficiently narrow foreground redshift distributions, and with intrinsic alignments and lens magnification controlled, these lensing ratios (LRs) become geometrical. Here we characterize the geometrical limit and develop an end-to-end optimized framework for the probe with LSST Y10 and Simons Observatory uncertainties. In our forecast for LSST Y10, the individual ratios are measured to a median precision of $1.0\%$, with a total signal-to-noise ratio of $360$. We then study the complementarity of LRs as an additional geometrical probe to constrain dark energy and curvature. We find that LRs improve the $w_0$--$w_a$ figure of merit of the combined CMB, Type Ia supernova, and baryon acoustic oscillation (BAO) data by 15\%, after marginalizing over systematic uncertainties. As an independent geometric probe, they also enable competitive leave-one-out tests, in which one of the probes is dropped, and we find they are most effective at replacing the BAO: adding LRs to CMB+supernova raises the figure of merit from 33\% to 66\% of that of the full CMB+supernova+BAO combination. We also find LRs are especially valuable in extended model spaces constrained with minimal assumptions: in a seven-bin reconstruction of the dark-energy density, they reduce per-bin uncertainties by up to 16\% at $z>1$, and by up to 50\% when freeing curvature and dropping the non-geometrical part of the CMB likelihood (i.e. keeping the acoustic angular scale part),  where they are able to break the degeneracy between curvature and dark energy.

\end{abstract}


\maketitle

\section{Introduction}\label{sec:intro}

Recent data challenge the cosmological constant as the description of cosmic acceleration. The combination of baryon acoustic oscillation (BAO), cosmic microwave background (CMB), and Type Ia supernovae (SNe) data prefers an evolving dark-energy density at about $3\sigma$ \citep{DESIY2BAO2025, Popovic2025, DESY6w0wa2026, Lee2026}. The preference is persistent but not yet decisive: the hints are still there when removing the SNe, the CMB or the BAO, but the constraining power weakens significantly. Also, the two distance probes depend on different assumptions: the SNe constrain relative distances, which are sensitive to the photometric cross-calibration between low- and high-redshift samples  \citep{Popovic2025}, while the BAO constrain distances in units of the sound horizon, whose absolute size is set by early-Universe physics when they are combined with the CMB. Some  degeneracies also affect the interpretation of the data, in particular among dark-energy evolution, spatial curvature and the neutrino masses \citep{Hannestad2005, Clarkson2007, ChenZaldarriaga2025}, yet for lack of constraining power these extensions are almost always analyzed separately or in pairs \citep[e.g.][]{Lodha2025}. Thus, it would be highly valuable to add an independent  geometrical probe to the cosmological toolkit, which would serve two purposes: enabling more competitive leave-one-out robustness tests and tightening constraints, especially on extended parameter spaces.

Lensing ratios (LRs) offer precisely this \citep{JainTaylor2003, BernsteinJain2004, Bernstein2006, Taylor2007}. The lensing signal around foreground galaxies factorizes into the projected mass distribution of the lenses and a purely geometric factor built from the observer--lens--source distances. If the same lens sample is observed with several background source planes, the mass factor is common to all of them and cancels in ratios of the cross-correlations, leaving ratios of angular diameter distances in the limit where the lens redshift distribution is infinitely narrow. In this geometrical regime the LRs therefore require no galaxy-bias model, are insensitive to the nonlinear matter power spectrum, and can use small angular scales that standard weak-lensing analyses must discard because of nonlinear and baryonic modeling uncertainties. They are also computationally cheap, since predicting distance ratios requires no nonlinear structure modeling. 

Different classes of ratios follow from the choice of source planes: galaxy samples only, or including the CMB \citep{Hu2007, DasSpergel2009}. The CMB source plane is particularly useful: its redshift is precisely known and it gives the longest available distance. On the other hand, galaxy-only ratios, with all source planes at low redshift, are less sensitive to cosmology and more dependent on photometric-redshift and shear-calibration uncertainties and in DES they have  proven useful as calibration and nuisance-mitigation tools \citep{y1-ggl, y3-ratios, y6-ggl}.

CMB lensing ratios have been measured three times before on data, all with spectroscopic lenses except the last. \citet{Miyatake2017} measured a single ratio at $17\%$ precision, combining CMASS lens galaxies with CFHTLenS galaxy lensing and \textit{Planck} CMB lensing, and \citet{Singh2016} measured one at $11\%$ with BOSS lenses, SDSS galaxy lensing and \textit{Planck}. \citetalias{PratBaxter2019}~\citep{PratBaxter2019} constrained six ratios at the $13$--$23\%$ level, with $10\%$ precision on their combined amplitude, extending the method to photometric lens galaxies with DES, SPT, and \textit{Planck} data. 


Forecasts have shown that future surveys can reach much higher precision. \citet{BermejoCliment2020} forecast CMB lensing ratios with \textit{Euclid}-like sources and spectroscopic lenses from \textit{Euclid}, DESI and SPHEREx for current and future CMB experiments, without systematic uncertainties, and \citet{Li2026} recently proposed kinematic lensing ratios, which use intrinsic galaxy shapes inferred from kinematics to suppress shape noise and most first-order lensing systematics, and forecast them for \textit{Roman}. 

In this work we develop and validate the end-to-end framework needed to bring geometric CMB lensing ratios to per-cent precision in the LSST $\times$ Simons Observatory era, including the systematic uncertainties of a realistic analysis, and we forecast their impact for the two purposes above.
We characterize the regime of validity of the geometric limit, construct tomographic lens and source samples optimized for the probe, with a lens selection that keeps the ratios geometric where lens magnification could otherwise bias them,  and introduce an estimator that shows negligible bias and approximately Gaussian sampling distributions. We also test spectroscopic lens samples from 4MOST, DESI and \textit{Euclid}.  Then,  we run simulated joint analyses with CMB, BAO, and SNe likelihoods across several extended cosmological models, including evolving dark energy, free curvature and free neutrino masses. We also test a non-parametric dark-energy model. Finally we quantify the cost of marginalizing over photometric-redshift and shear-calibration, and  the impact of intrinsic-alignments.

The paper is organized as follows. Section~\ref{sec:theory} presents the theory of lensing ratios and the limits of the geometric approximation. Section~\ref{sec:data} describes the simulated data: the LSST and Simons Observatory configurations, the lens and source samples, the covariance, the ratio estimator, the geometric-validity selection, and the external likelihoods. Section~\ref{sec:results} presents the cosmological results. 

Section~\ref{sec:conclusions} summarizes our findings. The appendices validate the estimator on noisy data and compare it with the DES Y3 shear-ratio estimator, present the ratios obtained with the spectroscopic lens sample from 4MOST, DESI and Euclid, decompose the lensing-ratio error budget, quantify the impact of photometric-redshift, shear-calibration, intrinsic-alignment and CMB-lensing systematics, and assess the source-sample optimization impact.

\section{Theory}\label{sec:theory}

\subsection{Galaxy-lensing cross-correlations}\label{sec:xcorr}

Gravitational lensing by foreground structure produces a coherent correlation between lens galaxy positions and background convergence fields. This work is built on taking ratios of two such cross-correlations:
\begin{enumerate}
  \item $\xi_{g\kappa_g}^{(ij)}(\theta)$: the galaxy--convergence correlation around lens galaxies in bin $i$ and source galaxies in  bin $j$. 
  \item $\xi_{g\kappa_{\rm CMB}}^{(i)}(\theta)$: the galaxy--CMB lensing convergence around the same lens galaxies in tomographic bin $i$.
\end{enumerate}
Under the Limber approximation \citep{Limber1953,Kaiser1992}, their corresponding angular cross-spectra at multipole $\ell$ reduce  to the line-of-sight integral
\be
  C_\ell^{g\kappa,ij} =
  \int_0^{\chi_H} \frac{d\chi}{f_K^2(\chi)}\,
  W_g^i(\chi)\, W_\kappa^j(\chi)\,
  P_m\!\left(\frac{\ell+\tfrac{1}{2}}{f_K(\chi)},z(\chi)\right),
  \label{eq:cl}
\ee
where $\chi$ is the comoving radial distance to redshift $z$, $\chi_H$ the comoving distance to the horizon, $P_m(k,z)$ is the nonlinear matter power spectrum at wavenumber $k$, evaluated at $k=(\ell+1/2)/f_K(\chi)$, $W_g^i(\chi)=b_i\,n_i^l(z)\,dz/d\chi$ is the galaxy position kernel with linear galaxy bias $b_i$ and normalized redshift distribution $n_i^l(z)$ of lens bin $i$, and $W_\kappa^j$ is the lensing kernel. $f_K(\chi)$ is the comoving transverse distance: $f_K(\chi)=\chi$ for a flat universe, $f_K(\chi)=|\Omega_k|^{-1/2}(c/H_0)\sinh\!\left(|\Omega_k|^{1/2}H_0\chi/c\right)$ for open ($\Omega_k>0$), and $f_K(\chi)=|\Omega_k|^{-1/2}(c/H_0)\sin\!\left(|\Omega_k|^{1/2}H_0\chi/c\right)$ for closed ($\Omega_k<0$).

For galaxy source bin $j$, with normalized redshift distribution $n_j^s(z)$, and for the CMB source plane respectively, the convergence kernels are
\bea
 & W_{\kappa_g}^j(\chi) =
  \frac{3H_0^2\Omega_m}{2c^2}\frac{f_K(\chi)}{a(\chi)}
  \int_\chi^{\chi_H} n_j^s(z')\frac{dz'}{d\chi'}
  \frac{f_K(\chi'-\chi)}{f_K(\chi')}\,d\chi',
  \label{eq:Wkg}\\[4pt]
 & W_{\kappa_{\rm CMB}}(\chi) =
  \frac{3H_0^2\Omega_m}{2c^2}\frac{f_K(\chi)}{a(\chi)}
  \frac{f_K(\chi_{\rm CMB}-\chi)}{f_K(\chi_{\rm CMB})},
  \label{eq:Wcmb}
\eea
where $a(\chi)$ is the scale factor, and with the CMB source plane at $z_{\rm CMB}=1089.80\pm0.21$ \citep{PlanckParams2018}. The corresponding real-space correlations are obtained through the spin-0$\times$spin-0 transform\footnote{We use convergence correlations rather than the usual tangential shear $\gamma_t$  for galaxy-galaxy lensing in order to keep the same scalar angular kernel for galaxy-source and galaxy-CMB lensing correlations. This is important because scale-independent ratios in real space require the same angular transform in numerator and denominator.}
\be
  \xi_{g\kappa}^{(ij)}(\theta) =
  \frac{1}{4\pi}\sum_\ell (2\ell+1)\,
  C_\ell^{g\kappa,ij}\,P_\ell(\cos\theta),
  \label{eq:hankel}
\ee
where $\theta$ is the angular separation and $P_\ell$ are the Legendre polynomials of degree $\ell$.

For the galaxy--CMB lensing cross-correlations, we mimic the properties of the reconstructed CMB convergence map before transforming to real space, following the procedure of \citet{Zhang2022}. We multiply the angular cross-spectra $C_\ell^{g\kappa_{\rm CMB},i}$ by a Gaussian beam with $1~{\rm arcmin}$ full-width-half-maximum, approximating the size of the SO beam, and restrict the harmonic transform of Eq.~\eqref{eq:hankel} to the multipole range $30 \le \ell \le 3000$ \citep[following][]{Ade2019}, the scales over which the SO lensing map is reliably reconstructed. The same beam and multipole range are applied consistently to numerator and denominator so that the ratios remain scale-independent. 

\subsection{Geometric lensing ratios}\label{sec:ratios} 

The lensing ratios are geometrical in the limit of an infinitely narrow lens redshift distribution. In this limit, the lensing cross-correlation expressions from the section above factorise into a term describing the projected lens mass distribution and a term depending only on distances:
\bea
  \xi_{g\kappa_g}^{(ij)}(\theta) &=&
  \Sigma^{(i)}(\theta)\,
  \langle\Sigma_{\rm cr}^{-1}\rangle^{(ij)},
  \label{eq:gkappa_gal}\\[4pt]
  \xi_{g\kappa_{\rm CMB}}^{(i)}(\theta) &=&
  \Sigma^{(i)}(\theta)\,
  \langle\Sigma_{\rm cr}^{-1}\rangle^{(i,{\rm CMB})}.
  \label{eq:gkappa_cmb}
\eea
$\Sigma^{(i)}(\theta)$ is the projected surface mass-density contrast around the lens galaxies and it carries all the lens properties: galaxy bias, matter clustering, and nonlinear structure. With tangential-shear correlations the corresponding mass factor would be the difference $\langle\Sigma(<\theta)\rangle-\Sigma(\theta)$; the convergence correlations used here measure the projected density profile itself. On the other hand the effective inverse critical surface densities $\langle\Sigma_{\rm cr}^{-1}\rangle$ are geometric factors. For a population of lens and source galaxies,
\be
  \langle\Sigma_{\rm cr}^{-1}\rangle^{(ij)}
  =
  \int dz_l\, n_i^l(z_l)
  \int dz_s\, n_j^s(z_s)\,
  \Sigma_{\rm cr}^{-1}(z_l,z_s),
  \label{eq:sigcr_eff}
\ee
and for the CMB source plane,
\be
  \langle\Sigma_{\rm cr}^{-1}\rangle^{(i,{\rm CMB})}
  =
  \int dz_l\, n_i^l(z_l)\,
  \Sigma_{\rm cr}^{-1}(z_l,z_{\rm CMB}).
  \label{eq:sigcr_cmb}
\ee
For a single lens-source pair, the inverse critical surface density is
\be
  \Sigma_{\rm cr}^{-1}(z_l,z_s)
  =
  \frac{4\pi G}{c^2}\,
  \frac{D_A(z_l)\,D_A(z_l,z_s)}{D_A(z_s)},
  \label{eq:sigcr}
\ee
where $D_A(z_l)$ and $D_A(z_l,z_s)$ are the angular diameter distance to the lens and between lens and source, and $\Sigma_{\rm cr}^{-1}(z_l,z_s)=0$  for $z_s\leq z_l$.

Then, for a fixed lens bin $i$ and two source galaxy bins $j\neq j'$, the galaxy--galaxy lensing ratio becomes
\be
  \mathcal{R}^{\rm GGL}_{(i,j,j')}
  =
  \frac{\xi_{g\kappa_g}^{(ij)}(\theta)}
       {\xi_{g\kappa_g}^{(ij')}(\theta)}
  =
  \frac{\langle\Sigma_{\rm cr}^{-1}\rangle^{(ij)}}
       {\langle\Sigma_{\rm cr}^{-1}\rangle^{(ij')}},
  \label{eq:ratio_ggl}
\ee
and for a lens bin $i$ and source galaxy bin $j$, the CMB lensing ratio is
\be
  \mathcal{R}^{\rm CMB}_{(i,j)}
  =
  \frac{\xi_{g\kappa_{\rm CMB}}^{(i)}(\theta)}
       {\xi_{g\kappa_g}^{(ij)}(\theta)}
  =
  \frac{\langle\Sigma_{\rm cr}^{-1}\rangle^{(i,{\rm CMB})}}
       {\langle\Sigma_{\rm cr}^{-1}\rangle^{(ij)}}.
  \label{eq:ratio_cmb}
\ee
Note we always keep the CMB in the numerator since it is typically the quantity with less signal-to-noise. Spelled out with the tomographic redshift distributions through Eqs.~\eqref{eq:sigcr_eff}--\eqref{eq:sigcr}, the geometric model can be written as:
\be
  \mathcal{R}^{\rm CMB}_{(i,j)}
  =
  \frac{\displaystyle\int dz_l\, n_i^l(z_l)\,
        D_A(z_l)\,\frac{D_A(z_l,z_{\rm CMB})}{D_A(z_{\rm CMB})}}
       {\displaystyle\int dz_l\, n_i^l(z_l)
        \int dz_s\, n_j^s(z_s)\,
        D_A(z_l)\,\frac{D_A(z_l,z_s)}{D_A(z_s)}},
  \label{eq:ratio_cmb_full}
\ee
and analogously for $\mathcal{R}^{\rm GGL}$, with the numerator replaced by a second source-averaged integral. Note the model is scale-independent so each bin combination is predicted by a single number, matched on the data side by compressing the measured correlations into one amplitude per combination (Sec.~\ref{sec:estimator}).

Considering systematic uncertainties, photo-$z$ mean-shift parameters $\Delta z_b$, one per tomographic bin, displace the input redshift distributions, $n_b(z) \to \tilde n_b(z) \propto n_b(z - \Delta z_b)$. Multiplicative shear calibration $m_j$, one per source bin, rescales the measured $\xi_{g\kappa_g}^{(ij)}$ by $(1+m_j)$; since the shear enters the denominator of the CMB ratios, the full model for cosmological parameters $\mathbf{p}$ becomes
\be
  \mathcal{R}^{\rm CMB,\,model}_{(i,j)}(\mathbf{p}, \Delta z_b, m_j)
  =
  \frac{1}{1+m_j}\,
  \frac{\langle\Sigma_{\rm cr}^{-1}\rangle^{(i,{\rm CMB})}\big[\tilde n^l_i\big]}
       {\langle\Sigma_{\rm cr}^{-1}\rangle^{(ij)}\big[\tilde n^l_i,\, \tilde n^s_j\big]},
  \label{eq:ratio_model_sys}
\ee
and the galaxy--galaxy lensing ratios of Eq.~\eqref{eq:ratio_ggl} analogously acquire a factor $(1+m_j)/(1+m_{j'})$. For the spectroscopic lens sample the lens redshifts are exact and only the source shifts remain. The priors on these parameters and their impact are quantified in Appendix~\ref{app:systematics}. 
The likelihood is Gaussian in the compressed ratios,
\be
  -2\ln\mathcal{L}
  =
  \left(\hat{\mathbf r} - \mathbf r_{\rm model}\right)^\top
  \hat{\mathbf{C}}_{\mathbf r}^{-1}
  \left(\hat{\mathbf r} - \mathbf r_{\rm model}\right),
  \label{eq:ratio_like}
\ee
with the full data vector $\hat{\mathbf r}$ and its  covariance $\hat{\mathbf{C}}_{\mathbf r}$ constructed in Secs.~\ref{sec:estimator} and \ref{sec:lr_cov}.

\subsection{Beyond the ideal geometric limit}\label{sec:geometry_breaking}

\begin{figure}
  \centering
  \includegraphics[width=0.5\textwidth]{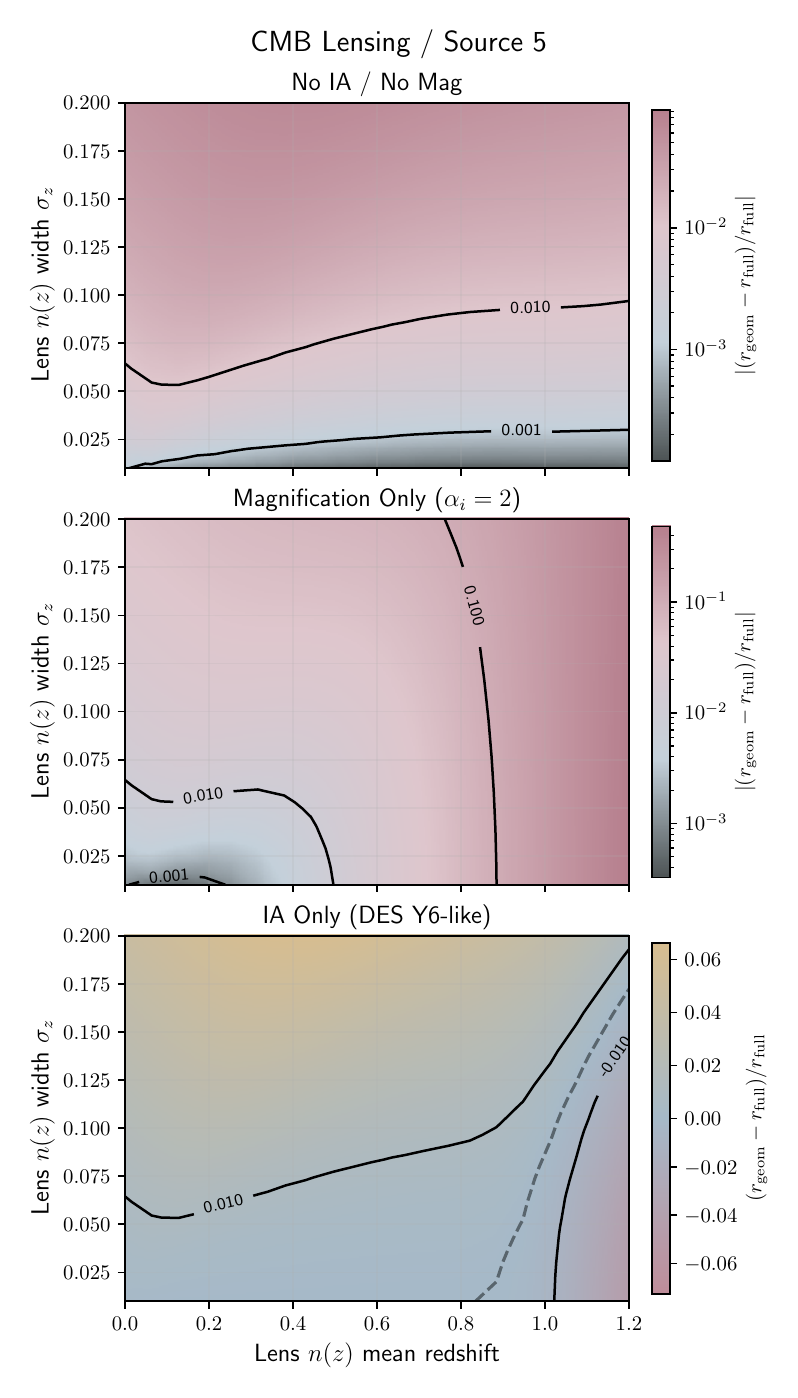}
  \caption{\textbf{Validity of the geometry-only approximation.}
Fractional difference between the geometry-only prediction and the full model for the CMB-lensing/source~5 ratio,
$\Delta r=(r_{\rm geom}-r_{\rm full})/r_{\rm full}$,
as a function of the mean redshift and width of a Gaussian lens redshift distribution.
The top panel includes neither lens magnification nor intrinsic alignments; the middle panel includes lens magnification only, with $\alpha_l=2$; and the bottom panel includes intrinsic alignments only, using the fiducial TATT-like IA parameters.
The top and middle panels show $|\Delta r|$ on logarithmic colour scales, while the bottom panel shows the signed residual on a linear colour scale to expose cancellations and zero crossings.
Each panel has its own colorbar. Solid contours mark representative residual levels, and the dashed contour in the bottom panel marks $\Delta r=0$.
}
  \label{fig:geometry_grid}
\end{figure}

The ratios in Eqs.~\eqref{eq:ratio_ggl} and \eqref{eq:ratio_cmb} are purely geometric only if the same lens-plane contribution factors out of numerator and denominator. In realistic photometric surveys this factorisation is approximate. The leading effects that can spoil the geometry-only limit are broad lens redshift distributions, magnification of the lens sample, and intrinsic alignments of source galaxies. 

First, if the lens redshift distribution is broad, the distance-dependent part of the signal cannot be pulled out of the line-of-sight integral of Eq.~\eqref{eq:cl}: the observable becomes an average over lens redshift, weighted by the galaxy bias, growth, and matter power spectrum across the lens distribution, and the ratio of two such integrals no longer reduces to a ratio of $\langle\Sigma_{\rm cr}^{-1}\rangle$ factors. The cancellation is a good approximation when the lens distribution is narrow enough that these terms vary slowly across it.

Second, the observed lens galaxy density is not only the intrinsic galaxy overdensity. It also receives a contribution from magnification bias \citep{Unruh2019, ElvinPoole2023, DESY6Magnification2026} because foreground structure changes both the apparent fluxes and observed solid angle of the lens sample:
\be
  \delta_g^{\rm obs}
  =
  \delta_g^{\rm int}
  +
  2(\alpha_l-1)\kappa_l,
  \label{eq:lens_magnification_delta}
\ee
where $\alpha_l$ is the logarithmic number-count slope of the lens sample and $\kappa_l$ is the convergence to the lens population. Substituting this observed density into the cross-correlation adds a magnification contribution,
\be
  C_\ell^{g_{\rm obs}\kappa_s}
  =
  C_\ell^{g_{\rm int}\kappa_s}
  +
  2(\alpha_l-1)\,
  C_\ell^{\kappa_l\kappa_s}.
  \label{eq:lens_magnification_cl}
\ee
The first term is the desired correlation between the lens density and the background convergence field. The second term is sourced by matter in front of the lens sample and has a different redshift kernel from the intrinsic lens-density term. It therefore does not share the same factorised lens-plane dependence, and its contribution does not cancel in ratios. This effect is especially important for broad or high-redshift lens samples, and for lens selections with number-count slopes far from $\alpha_l=1$.

Third, the source shape field is not a pure gravitational-lensing field. The observed source convergence or shear contains a lensing contribution and an intrinsic-alignment contribution,
\be
  \kappa_{\rm obs}^j
  =
  \kappa_{\rm G}^j
  +
  \kappa_{\rm IA}^j,
  \label{eq:obs_kappa_ia}
\ee
so that
\be
  C_\ell^{g\kappa_{\rm obs},ij}
  =
  C_\ell^{g\kappa_{\rm G},ij}
  +
  C_\ell^{g\kappa_{\rm IA},ij}.
  \label{eq:ia_cl}
\ee
The IA term traces correlations between lens galaxies and the intrinsic orientations of source galaxies. It does not depend on the same lensing-efficiency kernel and it appears  when the lens and source redshift distributions overlap. It therefore introduces both redshift and scale dependence that is not captured by the critical-surface-density factors in Eqs.~\eqref{eq:ratio_ggl} and \eqref{eq:ratio_cmb}.

\subsection{Where is the geometry-only approximation valid?}
\label{sec:geometry_grid}

We quantify the validity of the geometry-only approximation using a grid of Gaussian lens redshift distributions with varying mean $\bar z_l$ and width $\sigma_z$, keeping the galaxy-source distributions fixed to the fiducial LSST Y1 bins (see \S\ref{sec:surveys}). For each configuration, we compare the full prediction from \textsc{CosmoSIS} \citep{Zuntz2015CosmoSIS}, including the contributions described in Sec.~\ref{sec:xcorr}, with the geometry-only prediction obtained from the critical-surface-density factors in Eq.~(\ref{eq:ratio_cmb_full}) and its galaxy-galaxy lensing analogue. Figure~\ref{fig:geometry_grid} shows the fractional residual $(r_{\rm geom}-r_{\rm full})/r_{\rm full}$ for CMB-lensing/source~5 ratios.

The three panels show the baseline without intrinsic alignments or lens magnification, magnification only, and intrinsic alignments only. We adopt $\alpha_i=2$ for the magnification test, with $\alpha_i=1$ corresponding to no magnification, and the fiducial TATT parameters \citep{Blazek2019} $A_1=0.54$ and $A_2=-0.89$, together with their adopted redshift dependence (see \S\ref{app:systematics}), for the IA test. These choices provide illustrative contamination levels rather than a forecast for a specific lens selection. The first two panels show absolute residuals, whereas the IA panel retains their sign to expose cancellations.

Narrow lens distributions provide the most reliable geometrical ratios: increasing the width introduces different redshift weightings of the structure-dependent contributions in the numerator and denominator. For CMB-lensing/source~5, the baseline departures are generally larger at low lens redshift for a fixed width. Magnification instead produces a strong increase in the residual toward high lens redshift, where lens magnification contributions are more significant. Intrinsic alignments introduce a more structured response, including zero crossings where contributions of opposite sign cancel. These cancellations depend on the assumed IA parameters and shift when they change, so a small residual near a zero crossing does not indicate that a configuration is robust to IA contamination. These results motivate narrow lens selections with controlled magnification and IA contributions, with the acceptable width depending on redshift and the required modelling accuracy (we will do this in detail in \S\ref{sec:sim_ratios_cov}).

\subsection{Cosmological sensitivity of geometrical ratios}
\label{sec:cosmology_sensitivity}

We next use the same setup to identify which lens--source configurations respond most strongly to cosmology. We fix the far source to the CMB at $z_*=1090$ and scan the mean redshifts of Gaussian lens and galaxy-source distributions, holding their widths fixed at $\sigma_{z,l}=0.003$ and $\sigma_{z,s}=0.05$. On a $25\times25$ grid spanning $0.05\leq\bar z_l\leq1.2$ and $0.1\leq\bar z_s\leq2.5$, we evaluate
\begin{equation}
S_p \equiv \frac{\partial\ln r_{\rm geom}}{\partial p}
\simeq
\frac{\ln r_{\rm geom}(p+\Delta p)
      -\ln r_{\rm geom}(p-\Delta p)}
     {2\Delta p},
\end{equation}
for $p\in\{\Omega_m,\Omega_k,w,w_a\}$, where $w(a)=w+w_a(1-a)$ \citep{ChevallierPolarski2001, Linder2003}. This logarithmic derivative measures the fractional response of the ratio per unit change in the parameter.

\begin{figure}
  \centering
  \includegraphics[width=0.9\linewidth]{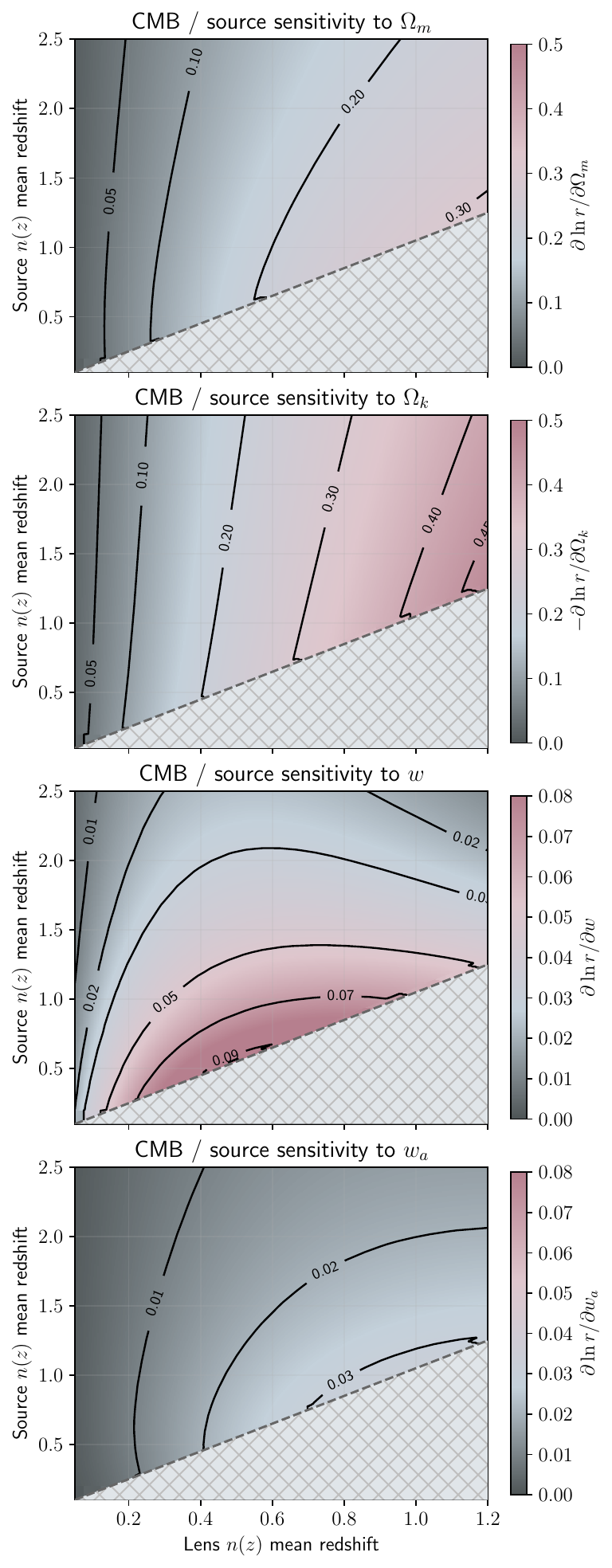}
  \caption{
  \textbf{Cosmological sensitivity of geometrical lensing ratios.}
  Fractional response of the geometry-only CMB-lensing/galaxy-source
  ratio, $S_p=\partial\ln r_{\rm geom}/\partial p$, as a function of
  the mean lens and galaxy-source redshifts. The lens and source
  distributions are Gaussian with fixed widths
  $\sigma_{z,l}=0.003$ and $\sigma_{z,s}=0.05$, respectively,
  and the CMB source plane is fixed at $z_*=1090$.
  The panels show the responses to $\Omega_m$, $\Omega_k$, $w$,
  and $w_a$. For curvature we display $-S_{\Omega_k}$, since its
  response is negative throughout the plotted domain.
  Colours show finite-difference derivatives
  interpolated from the sampled grid.
  }
  \label{fig:cosmology_sensitivity}
\end{figure}

The derivatives are evaluated around $\Omega_m=0.3$, $\Omega_k=0$, $w=-1$, and $w_a=0$, with $H_0=70\,{\rm km\,s^{-1}\,Mpc^{-1}}$ and $\sum m_\nu=0.06\,{\rm eV}$, using steps $\Delta\Omega_m=\Delta\Omega_k=0.02$, $\Delta w=0.05$, and $\Delta w_a=0.10$. We vary each parameter separately, holding the remaining independent
parameters fixed. When varying $\Omega_m$ or $\Omega_k$, we adjust
the present-day dark-energy density so that the density parameters,
including curvature and radiation, sum to unity. When varying $w$
or $w_a$, the present-day densities remain fixed. The sampled derivatives are interpolated for visualization, and configurations with $\bar z_s-\bar z_l\leq0.05$ are excluded.

Figure~\ref{fig:cosmology_sensitivity} shows the responses to $\Omega_m$, $\Omega_k$, $w$ and $w_a$. The curvature derivative is negative throughout the displayed domain, so we plot $-S_{\Omega_k}$ to compare its magnitude directly with the positive responses of the other parameters. The colour scales span $0$--$0.5$ in the upper 2 panels and $0$--$0.08$ in the two lower ones, allowing comparison between those pairs of parameters (larger responses saturate at the upper colour limit).

The matter-density response and the magnitude of the curvature response generally increase toward higher lens redshift. In contrast, the response to $w$ favours lower-to-intermediate lens redshifts, while that to $w_a$ extends to higher redshift. Because each parameter's response peaks in a different region of the lens–source plane, a single ratio constrains only one degenerate combination of parameters, whereas ratios spanning a wide range of lens redshifts weight these responses differently and therefore constrain independent combinations. Also, a galaxy-source distribution relatively close behind the lenses generally enhances the fractional response, although the dependence on source redshift is weaker for curvature. Across the scanned domain, the maximum response magnitude is larger for curvature than for matter density\footnote{The large curvature response has a purely geometric origin, pointed out by \citet{Bernstein2006}. The lensing efficiency depends on all three sides of the observer--lens--source triangle. In a flat universe the lens--source distance follows directly from the observer--lens and observer--source distances, but with curvature it acquires an additional term proportional to $\Omega_k$. Curvature therefore changes the ratios even when the distance--redshift relation is held fixed. In principle, lensing around a common lens sample with several source planes can thus solve simultaneously for the distance--redshift relation, treated as a free function, and the curvature, relying only on the geometry of the Robertson--Walker metric and not on a model for the expansion history, i.e.\ for dark energy or the growth of structure.}, and larger for $w$ than for $w_a$.

The optimal sample must therefore balance cosmological sensitivity against measurement noise and the modelling requirements identified in Sec.~\ref{sec:geometry_grid}. In particular, nearby lens and source distributions can have appreciable overlap and weak lensing efficiency despite a large fractional response. Parameter degeneracies, redshift uncertainties, and the covariance between ratios must also enter a forecast of the achievable constraints. We use the validity and sensitivity tests together to guide the COSMOS-based sample selection described in the next section.

\section{Simulated Data Setup}\label{sec:data}

\begin{table}
    \renewcommand{\arraystretch}{1.15}
    \centering
    \caption{\textbf{Fiducial cosmological parameters and prior ranges.} Square brackets denote flat priors.   $\sigma_8$ is the value derived from the fixed $A_s$ at the fiducial
    cosmology with $\sum m_\nu = 0.06$\,eV.
    $^\dagger\Omega_b$ is varied with a flat prior in all chains that include BAO,
    $\theta_*$, or the compressed CMB prior, where it enters the sound-horizon
    computation; it is held fixed in LR-only and SN-only chains, where it has
    negligible effect on the purely geometric observables.
    }
    \label{tab:fiducial}
    \begin{tabular}{lcc}
        \toprule
        & Fiducial & Prior \\
        \midrule
        \multicolumn{3}{l}{\textit{Cosmological}} \\
        $\Omega_m$         & 0.3156               & $[0.10,\;0.60]$             \\
        $h$                & 0.6727               & $[0.50,\;0.90]$             \\
        $\Omega_b$         & 0.0492               & $[0.03,\;0.07]^\dagger$     \\[3pt]
        \multicolumn{3}{l}{\textit{Model extensions (when applicable)}} \\
        $w$                & $-1$                 & $[-3.0,\;-\tfrac{1}{3}]$    \\
        $w_a$              & 0                    & $[-3.0,\;3.0]$              \\
        $\Omega_k$         & 0                    & $[-0.25,\;0.25]$            \\
        $\sum m_\nu$ [eV]  & 0.06           & $[0.06,\;0.6]$                         \\[3pt]
        \multicolumn{3}{l}{\textit{Fixed in all chains}} \\
        $n_s$              & 0.9645               & ---                         \\
        $A_s$              & $2.129\times10^{-9}$ & ---                         \\
        $\sigma_8$         & 0.819          & ---                         \\
        $\tau$             & 0.054                & ---                         \\
        \bottomrule
    \end{tabular}
\end{table}

\begin{figure*}
    \centering
    \includegraphics[width=\textwidth]{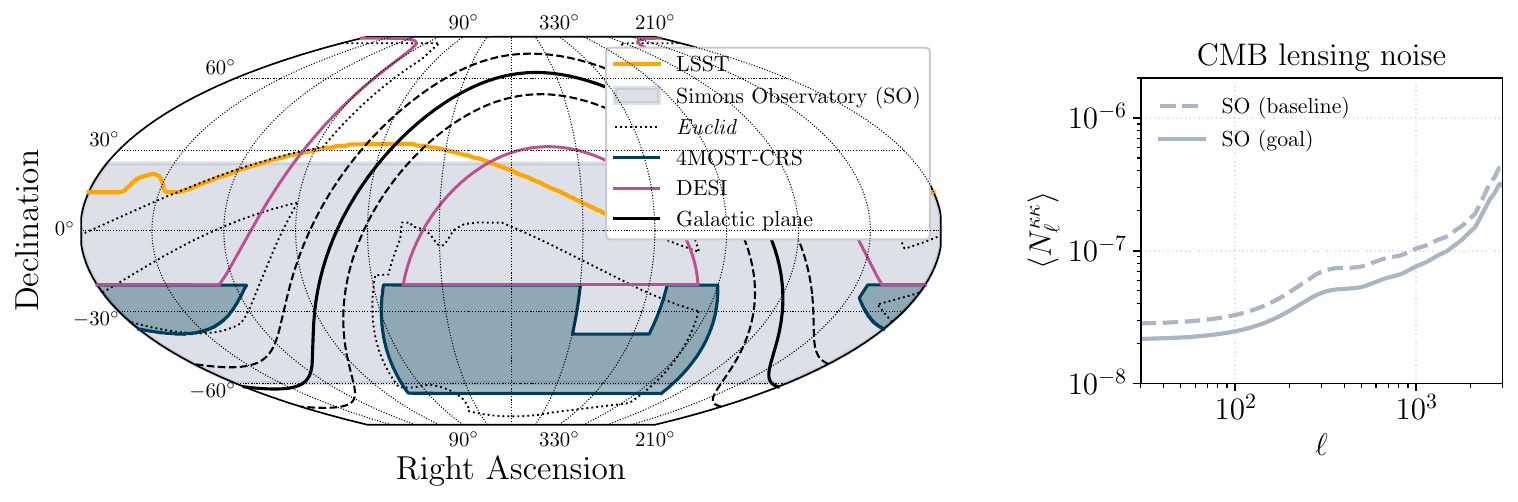}
  \caption{\textbf{Survey characteristics.} \textit{Left:} Footprints of the surveys used in this work: LSST and the Simons Observatory (SO), together with the spectroscopic surveys combined into the alternative lens sample of Appendix~\ref{app:lens_specz}: 4MOST-CRS, DESI, and \textit{Euclid}. \textit{Right:} CMB lensing noise power spectra. For SO we show both the baseline sensitivity (dashed), adopted for LSST Y1 cross-correlations ($\sim$2028), and the goal sensitivity (solid), adopted for LSST Y10. }
    \label{fig:surveys}
\end{figure*}

\begin{figure}
    \centering
    \includegraphics[width=0.5\textwidth]{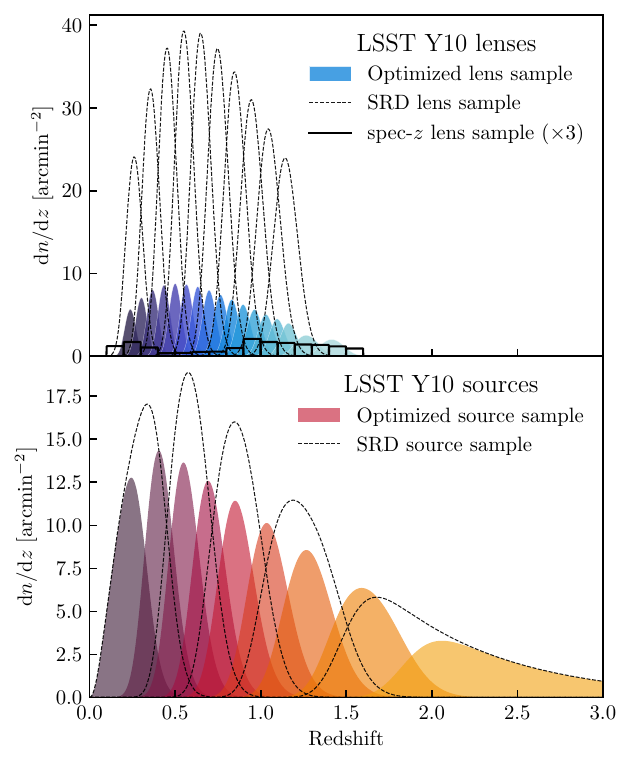}
  \caption{\small \textbf{Tomographic lens and source samples.} Number density per unit redshift of the LSST Y10 tomographic bins used in this work. Filled curves show the narrow lens (17 bins) and source (9 bins) samples optimized for lensing-ratio measurements; dashed black outlines show the corresponding standard LSST SRD Y10 samples (10 lens / 5 source bins). The solid black outline in the top panel shows the alternative combined 4MOST+DESI+\textit{Euclid} spectroscopic lens sample that we define and test in  Appendix~\ref{app:lens_specz}, with three times its number density for visibility.}
    \label{fig:nzs}
\end{figure}

All simulated data vectors and covariance matrices are computed at the fiducial cosmology listed in Table~\ref{tab:fiducial}. These values follow the LSST DESC Science Requirements Document \citep{Mandelbaum2018}, and are the same as those adopted by \citet{Zhang2022}, except that we additionally adopt the minimal neutrino mass $\sum m_\nu = 0.06\,\mathrm{eV}$.

\subsection{LSST}\label{sec:surveys}

The Vera C. Rubin Observatory has already begun full operations with its Legacy Survey of Space and Time (LSST) \citep{Ivezic2019}.  We forecast two LSST epochs. The fiducial configuration targets the completed ten-year survey (Y10): following the LSST DESC Science Requirements Document \citep{Ivezic2019, Mandelbaum2018}, we assume the Y10 gold galaxy sample ($i < 25.3$) with an effective weak-lensing source density of $27$~arcmin$^{-2}$, from which the tomographic samples of Sec.~\ref{sec:samples} are constructed, over the $14\,300$~deg$^2$ ($f_{\rm sky} = 0.347$) that overlap the Simons Observatory lensing map after galactic masking, see Fig.~\ref{fig:surveys} and Table~\ref{tab:cases_updated}). The Y1 configuration corresponds to the DR1 epoch (expected in 2028), with shallower samples ($10$~arcmin$^{-2}$ of sources over $12\,300$~deg$^2$), and is used to assess how much of the science case is already accessible with the first data release. We describe below how we split the galaxies in redshift bins.

\subsubsection{Lens and source optimized samples}\label{sec:samples}

The analysis of Sec.~\ref{sec:geometry_grid} motivates lens
samples with narrow redshift distributions spanning a broad
redshift range, together with control of lens magnification
and lens--source overlap. First, for LSST Y1, we investigate a practical route to
such selections using the COSMOS2020 catalog
\citep{Weaver_2022}. We train a $48\times48$ self-organizing map
(SOM), using the noise-aware algorithm of \citet{Sanchez_2020},
on six-band aperture fluxes and their uncertainties:
CFHT $u$ and HSC $grizy$, used as proxies for Rubin $ugrizy$.
We use an $i<23$ selection 
on photometry that has been degraded to an LSST Y1-like depth model, with
clean-footprint, galaxy-classification, and LePhare quality
cuts. The COSMOS LePhare redshift estimates provide an
empirical distribution $p(z\mid c)$ for each SOM cell $c$.
Tomographic samples are constructed from disjoint sets of
cells with similar mean redshifts and narrow individual
redshift distributions (cells that would broaden a bin or
introduce substantial tails are not assigned to any tomographic bins). The
resulting bin $n(z)$ is the galaxy-count-weighted mixture of
its constituent cell distributions.

The same SOM representation allows magnification response to
enter the sample selection alongside redshift and density, as it has been demonstrated recently by \citet{alarcon2026weaklensingbigmagnification}.
Following the flux-perturbation approach of
\citet{ElvinPoole2023,DESY6Magnification2026}, we perturb all
band fluxes by $f_b\rightarrow(1+2\delta\kappa)f_b$, with
$\delta\kappa=0.01$, and repeat the assignment to the fixed
SOM. For a fixed set of selected cells, the fractional
change in its population estimates the selection response,
\begin{equation}
 C_{{\rm sample},i}
 \simeq
 \frac{N_i(\delta\kappa)-N_i(0)}
      {\delta\kappa\,N_i(0)}.
\end{equation}
Including the solid-angle dilution gives
$C_{{\rm total},i}=C_{{\rm sample},i}-2
=2(\alpha_{l,i}-1)$ in the convention of
Sec.~\ref{sec:geometry_breaking}. Thus, the relevant target
is $C_{{\rm sample},i}\simeq2$, rather than zero selection
response. Cells can be combined to approach this target
while maintaining a narrow $n(z)$ and sufficient density (individual cells are not required to have a null
response). We give greater priority to magnification
suppression at intermediate and high lens redshift, where
the foreground-lensing contribution is more important and
does not share the lens-plane dependence that cancels in
the geometric ratio. Selection strategies targeting this
cancellation are discussed further in \citet{alarcon2026weaklensingbigmagnification}. 

These tests motivate the need for joint redshift--magnification
selection to retain geometrical lensing ratios. A complete calibration
must also include objects crossing the parent selection
boundary, detection and size-dependent selection effects,
and uncertainty in the estimated response. For a
measurement, the full selection response would be
calibrated using source-injection such as \textsc{Balrog} in DES \citep{ElvinPoole2023,DESY6Magnification2026}, with a procedure analogous to that of \citep{alarcon2026weaklensingbigmagnification}. 
For this paper, the COSMOS exercise yields a candidate LSST Y1 lens
configuration with 13 narrow tomographic bins, spanning
mean redshifts $\bar z\simeq0.17$--$0.94$, with redshift
widths $\sigma_z\simeq0.02$--$0.06$ and estimated densities
of approximately $0.06$--$0.20$~arcmin$^{-2}$ per bin.
For the six highest-redshift bins
($\bar z\simeq0.65$--$0.94$), the estimated flux-only
responses satisfy $|C_{\rm total}|\lesssim0.3$, while
retaining narrow redshift distributions (see summary in Table~\ref{tab:cases_updated}). 

Motivated by this sample-design strategy, we adopt an
analytic LSST Y10 lens forecast with per-galaxy photo-$z$
scatter $\sigma_z=0.02(1+z)$, compared with
$0.03(1+z)$ for the standard SRD lens sample. We use
15 uniform photo-$z$ intervals over $0.2<z_{\rm phot}<1.2$,
supplemented by intervals $1.2<z_{\rm phot}<1.35$ and
$1.35<z_{\rm phot}<1.5$. The parent distribution is
normalized to $10$~arcmin$^{-2}$ in the first 15 bins,
and the densities of the two additional bins are reduced
to continue the declining high-redshift density trend.
This gives 17 bins with mean redshifts
$\bar z=0.24$--$1.42$ and a total density of
$10.8$~arcmin$^{-2}$ (Figure~\ref{fig:nzs}). This density was chosen conservatively compared to that of \citet{Zhang2022} (35.7 arcmin$^{-2}$), to leave a density budget for imposing a joint redshift-magnification selection on top of the photo-$z$ cuts. These densities,
redshift widths, and negligible magnification are adopted
forecast specifications, motivated by the COSMOS tests.

The sources are the SRD Y10 population repartitioned into
9 equal-number photometric-redshift bins, preserving the
total source $n(z)$ and effective density of
$27$~arcmin$^{-2}$.

As an alternative to the photometric
lenses, we assemble a spectroscopic sample combining 4MOST \citep{deJong2019}, DESI \citep{DESI2016}, and \textit{Euclid} \citep{EuclidMellier2025} (Appendix~\ref{app:lens_specz}). Spectroscopic redshifts
minimize the lens photo-$z$ uncertainty but do not remove
magnification, whose response must still be characterized, following the procedure described above, for the corresponding galaxy selections.

Table~\ref{tab:cases_updated} lists the specifications of
the forecast configurations. We adopt the optimized lens
and source samples of Figure~\ref{fig:nzs} as fiducial.
Appendix~\ref{app:source_opt} shows that standard SRD
source binning performs nearly as well in most
configurations. The principal sample-design requirement
is therefore the construction of suitably narrow lens
bins with controlled magnification, rather than a
dedicated source selection.

\subsection{Simons Observatory}\label{sec:so}

The Simons Observatory (SO) will provide high-precision CMB
lensing maps over a wide footprint overlapping with LSST
(Figure~\ref{fig:surveys}). We assume approximately
$20{,}000\,\mathrm{deg}^2$ of SO coverage, with roughly
$15{,}000\,\mathrm{deg}^2$ remaining after Galactic masking;
the analysis footprints are specified in Table~\ref{tab:cases_updated}.
The Advanced SO upgrade is designed to approximately double
the LAT mapping speed \citep{SimonsObservatory2025}.

We adopt public lensing-reconstruction noise power spectra
from \citet{Ade2019}. For LSST Y1 we use the
\textit{baseline} curve (SENS1)\footnote{\url{https://github.com/simonsobs/so_noise_models/blob/master/LAT_lensing_noise/lensing_v3_1_1/nlkk_v3_1_0_deproj0_SENS1_fsky0p4_it_lT30-3000_lP30-5000.dat}},
and for LSST Y10 we use the \textit{goal} curve
(SENS2)\footnote{\url{https://github.com/simonsobs/so_noise_models/blob/master/LAT_lensing_noise/lensing_v3_1_1/nlkk_v3_1_0_deproj0_SENS2_fsky0p4_it_lT30-3000_lP30-5000.dat}}.
These choices define our forecast configurations; the noise
achieved at each epoch will depend on the accumulated
observing time, survey coverage, and reconstruction method.
Both spectra correspond to minimum-variance (MV)
reconstruction with no foreground deprojection and iterative
lensing reconstruction \citep[see also][]{SimonsObservatory2025}.

\begin{table*}[ht]
\centering
\caption{\textbf{Simulation specifications for each setup.} $\bar{z}$ denotes the weighted mean redshift of a tomographic bin.  The LSST lens galaxy biases follow $b_i = b_0/G(\bar z_i)$, with $G$ the linear growth factor at the fiducial cosmology and $b_0 = 1.05$ (Y1) and $0.95$ (Y10), as in \citet{Zhang2022}; the spec-$z$ biases are tracer-specific (see Appendix~\ref{app:lens_specz}). Source number densities are split equally among all bins; shape noise is $\sigma_e = 0.26$ per ellipticity component in all cases. The Gaussian photo-$z$ mean-shift priors $\sigma(\Delta z_b)$ and shear-calibration priors $\sigma_m$ (per-bin ranges quoted here) are derived in Appendix~\ref{app:systematics}. }
\label{tab:cases_updated}
\begin{tabular}{lccc}
\toprule
 & \textbf{LSST Y1} & \textbf{LSST Y10} & \textbf{Spec-$z$} \\
\midrule
\multicolumn{4}{l}{\textit{Survey \& CMB setup}} \\
$f_\mathrm{sky}$            & 0.298 (12{,}300 deg$^2$) & 0.347 (14{,}300 deg$^2$) & 0.259 (10{,}700 deg$^2$) \\
SO noise level              & Baseline (SENS1) & Goal (SENS2) & Goal (SENS2) \\
\midrule
\multicolumn{4}{l}{\textit{Lens galaxy sample}} \\
Sample                       & Optimized  & Optimized  & 4MOST+DESI+\textit{Euclid}  \\
$N_\mathrm{bins}^\mathrm{lens}$              & 13 & 17 & 16 \\
$\bar{z}^\mathrm{lens}$ range                & 0.17--0.94 & 0.24--1.42 & 0.14--1.55 \\
$n_\mathrm{lens}^\mathrm{tot}$ [arcmin$^{-2}$] & 1.59 & 10.76 & 0.57 \\
$b_i$ range                                  & 1.15--1.68 & 1.08--1.86 & 1.34--2.08 \\
$\sigma(\Delta z_b)$ prior range       & 0.0058--0.0097 & 0.0059--0.0133 & 0  \\[4pt]
\midrule
\multicolumn{4}{l}{\textit{Source galaxy sample}} \\
$N_\mathrm{bins}^\mathrm{src}$     & 5 & \multicolumn{2}{c}{9} \\
$\bar{z}_i^\mathrm{src}$
    & 0.25, 0.46, 0.66, 0.92, 1.53
    & \multicolumn{2}{c}{0.24, 0.42, 0.56, 0.71, 0.87, 1.05, 1.29, 1.63, 2.36} \\
$n_i^\mathrm{src}$ [arcmin$^{-2}$]   & 2.0 (equal per bin); total 10.0 & \multicolumn{2}{c}{3.0 (equal per bin); total 27.0} \\
$\sigma(\Delta z_b)$ prior range & 0.0025--0.0051 & \multicolumn{2}{c}{0.0017--0.0045} \\
$\sigma_m$ prior                 & 0.013 & \multicolumn{2}{c}{0.0040} \\[4pt]
\midrule
\multicolumn{4}{l}{\textit{Scale cuts (real space)}} \\
$\theta$ range [arcmin]             & \multicolumn{3}{c}{$2.5$--$250$ (20 log-spaced bins)} \\
$\ell_\mathrm{max}$ (theory)        & \multicolumn{3}{c}{3000} \\
$\theta$ range used in the ratios & \multicolumn{3}{c}{$\theta \leq 100$ arcmin (16 of the 20 bins)} \\
\bottomrule
\end{tabular}
\end{table*}

\subsection{Covariance of the cross-correlations}\label{sec:covariance}

We compute an analytic Gaussian covariance matrix with
\textsc{CosmoSIS} \citep{Zuntz2015CosmoSIS}, which we have
extended to support CMB lensing cross-correlations within
the \texttt{save\_2pt} module. 
For Gaussian fields and in the $f_{\rm sky}$ approximation for the mask, different multipoles are uncorrelated, $\mathrm{Cov}(C_\ell, C_{\ell'}) \propto \delta_{\ell\ell'}$, with
\begin{equation}
  \mathrm{Cov}\!\left(C_\ell^{12,ij},\,C_\ell^{34,kl}\right)
  = \frac{\tilde{C}_\ell^{13,ik}\,\tilde{C}_\ell^{24,jl}
         + \tilde{C}_\ell^{14,il}\,\tilde{C}_\ell^{23,jk}}
         {f_{\mathrm{sky}}\,(2\ell+1)},
  \label{eq:cl_cov}
\end{equation}
where field labels $1,2,3,4\in\{\kappa_g,\kappa_{\rm CMB},g\}$
and $i,j,k,l$ denote tomographic bins. Here,
$\tilde{C}_\ell$ denotes the sum of the signal and noise
power spectra, and $f_{\mathrm{sky}}$ is the overlapping
sky fraction. Noise contributions are included in the
auto-spectra and assumed to vanish between distinct
fields or disjoint tomographic samples. The beam and multipole range of Sec.~\ref{sec:xcorr} are applied to the data vector but not to the covariance, which we compute from the unfiltered spectra.

To obtain the real-space covariance we propagate Eq.~\eqref{eq:cl_cov} through the $C_\ell$-to-$\xi$ transformation.  Using the full-sky discrete sum $\xi^{(p)}(\theta) = \sum_\ell F^{(p)}_{\theta,\ell}\,C_\ell$, where the filter matrix $F^{(p)}_{\theta,\ell}$ encodes Legendre polynomials for galaxy clustering ($p=00$), associated Legendre polynomials for galaxy--shear ($p=02+$), and spin-weighted functions for shear--shear ($p=22\pm$), the covariance block becomes
\begin{equation}
  \Sigma^{(pq)}_{\theta_a,\theta_b}
  = \sum_\ell F^{(p)}_{\theta_a,\ell}\;
    \mathrm{Cov}(C_\ell^{ij},C_\ell^{kl})\;
    F^{(q)}_{\theta_b,\ell}.
  \label{eq:xi_cov}
\end{equation}

The noise contribution is treated differently depending on whether $N_\ell$ is white (i.e. scale independent). For galaxy positions and shapes the noise is $\ell$-independent, so the noise--noise term in Eq.~\eqref{eq:xi_cov} is formally a $\delta$-function in angle; a sum truncated at $\ell_{\max}$ converges poorly to it. We therefore subtract this term in harmonic space and add its exact real-space form, which is diagonal in angle,
\begin{equation}
  \Sigma^{\rm NN}_{\theta_a\theta_b} = \delta_{ab}
  \frac{N^{13}N^{24} + N^{14}N^{23}}{8\pi^2 f_{\rm sky} \, \theta_a \Delta\theta_a}
\end{equation}
with an additional factor of two for $\xi_\pm$ from $E$- and $B$-modes, and where $\delta_{ab}$ restricts the term to the diagonal, $\theta_a = \theta_b$. This is the usual pair-counting variance. The remaining signal and signal--noise terms are obtained from the explicit sum over $\ell$ in Eq.~\eqref{eq:xi_cov}, evaluated on an upsampled $\theta$ grid and averaged into the data bins with weights $w_i = \theta_i\Delta\theta_i$.

The CMB lensing reconstruction noise $N_\ell^{\kappa_{\rm CMB}}$ is strongly $\ell$-dependent (Fig.~\ref{fig:surveys}), so its noise--noise term is not diagonal in $\theta$ and has no pair-counting analogue. For every block involving $\kappa_{\rm CMB}$ we therefore keep the full observed spectra $\tilde C_\ell = C_\ell + N_\ell$ in Eq.~\eqref{eq:xi_cov} and evaluate all terms, including noise--noise, through the same $\ell$ sum. As $\kappa_{\rm CMB}$ is spin-0, $\xi_{g\kappa}$ uses the $p=00$ (Legendre) filter.

Partial sky coverage enters the covariance solely through the effective sky fraction $f_{\rm sky}$ in Eq.~\eqref{eq:cl_cov}. We do not apply a mode-coupling matrix (e.g.\ via \textsc{NaMaster}; \citealt{Alonso2019}), so the mask is not deconvolved at the power-spectrum level; this $f_{\rm sky}$ approximation is expected to slightly overestimate the covariance relative to a full pseudo-$C_\ell$ treatment. 

We validate our covariance against the independent Gaussian analytic estimate of \citet{Zhang2022} for the galaxy$\times$CMB-lensing cross-correlation $\xi_{g\kappa}$, computed with \textsc{TJPCov} \citep{TJPCov}.
Comparing the diagonal variance $C_{ii}^{g\kappa}$ across all five lens bins, we find agreement at the $\lesssim 10\%$ level, with our values systematically larger; the two calculations differ in their treatment of the mask (\citealt{Zhang2022} use \textsc{NaMaster}, we use the $f_{\rm sky}$ approximation) and in details of the noise and binning.

\subsection{Lensing ratios estimator}\label{sec:estimator}

\begin{figure}
    \centering
    \includegraphics[width=\columnwidth]{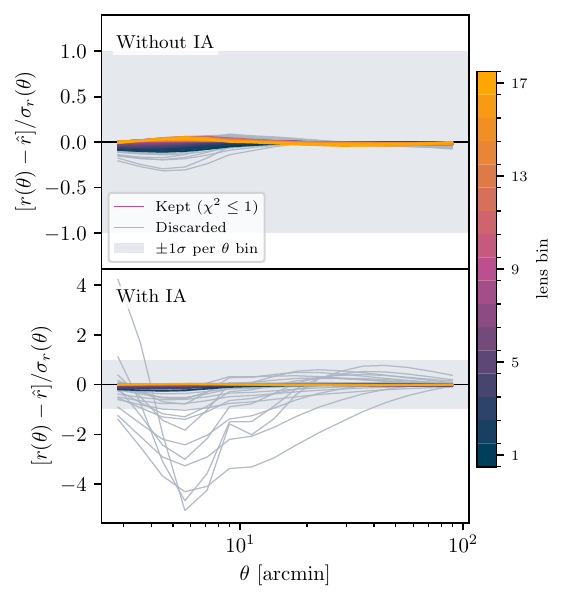}
    \caption{\textbf{Scale dependence of the lensing ratios.} Per-angular-bin CMB ratio $r(\theta) = \xi_{g\kappa_{\rm CMB}}(\theta)/\xi_{g\kappa}(\theta)$ of the noiseless LSST Y10 data vector relative to the compressed ratio $\hat r$ of the same lens--source combination, in units of the per-bin ratio uncertainty $\sigma_r(\theta)$, over the fiducial range $\theta \leq 100$~arcmin, for all 84  combinations. Combinations retained by the $\chi^2 \leq 1$ selection are colored by lens bin and discarded ones are shown in gray; the band marks $\pm1\sigma$. Top:  data vector without intrinsic alignment (IA) contamination. Bottom: data vector contaminated by IA and analyzed with its own selection.}
    \label{fig:ratio_scale_dependence}
\end{figure}

In the geometry-only limit, the ratios $\mathcal{R}^{\rm GGL}$ and $\mathcal{R}^{\rm CMB}$ defined in Eqs.~\eqref{eq:ratio_ggl}--\eqref{eq:ratio_cmb} are scale-independent. For realistic photometric lens bins, this scale independence is approximate, and we compress each ratio to a single scalar over a range of angular scales. Figure~\ref{fig:ratio_scale_dependence} shows that the residual scale dependence is small compared with the per-angular-bin uncertainties: the ratios, each measured to about $5\%$, stay within $0.3\sigma$ of the compressed value for every combination in the absence of intrinsic alignments. Intrinsic alignments introduce scale dependence of up to $5\sigma$ in some combinations; the $\chi^2$ selection of Sec.~\ref{sec:sim_ratios_cov} removes these, and the retained ratios stay within $0.25\sigma$.

In this section we describe the estimator for a single ratio combination. Section~\ref{sec:lr_cov} then assembles all lens--source combinations into the ratio data vector $\hat{\mathbf r}=(\hat r_1,\ldots,\hat r_N)$, where the quantities defined here carry an index $a$ when referring to combination $a$. For one combination, with numerator and denominator correlation functions $\boldsymbol{\xi}_{\rm num}$ and $\boldsymbol{\xi}_{\rm den}$ measured in $n_\theta$ angular bins within the selected range, we use an average-then-divide estimator motivated by the generalized-least-squares (GLS) fit of the amplitude $r$ of the linear model $\boldsymbol{\xi}_{\rm num}=r\,\boldsymbol{\xi}_{\rm den}$. The estimator is
\be
  \hat r =
  \frac{\mathbf v^\top\hat{\boldsymbol{\xi}}_{\rm num}}
       {\mathbf v^\top\hat{\boldsymbol{\xi}}_{\rm den}},
  \label{eq:ratio_estimator}
\ee
where hats denote measured quantities and $\mathbf v$ is the  weight vector, 
which we construct from the   covariance $\mathbf M$ of the residual of the fit $\boldsymbol{\xi}_{\rm num}-r\,\boldsymbol{\xi}_{\rm den}$. Starting from the joint $2n_\theta\times2n_\theta$ analytic covariance of the stacked vector $\mathbf d=(\boldsymbol{\xi}_{\rm num},\boldsymbol{\xi}_{\rm den})$, computed as in Sec.~\ref{sec:covariance}, we write
\be
  \mathbf C_{\mathbf d}\equiv
  \mathrm{Cov}\!
  \begin{pmatrix}
    \boldsymbol{\xi}_{\rm num}\\
    \boldsymbol{\xi}_{\rm den}
  \end{pmatrix}
  =
  \begin{pmatrix}
    \mathbf C_{\rm nn} & \mathbf C_{\rm nd}\\
    \mathbf C_{\rm dn} & \mathbf C_{\rm dd}
  \end{pmatrix}.
  \label{eq:ratio_cov_blocks}
\ee
Fixing the ratio in each angular bin at its fiducial cosmology value,
$r_k^{\rm fid}=\xi_{\rm num}^{\rm fid}(\theta_k)/
\xi_{\rm den}^{\rm fid}(\theta_k)$,
the residual is a linear transformation of the data,
\be
  \begin{aligned}
    \boldsymbol{\xi}_{\rm num}
      -\mathbf R\,\boldsymbol{\xi}_{\rm den}
      &=\mathbf A\,\mathbf d,\\
    \mathbf A&=
      \begin{pmatrix}\mathbf I & -\mathbf R\end{pmatrix},
    \qquad
    \mathbf R=\mathrm{diag}
      \big(r_1^{\rm fid},\ldots,r_{n_\theta}^{\rm fid}\big),
  \end{aligned}
  \label{eq:residual_linear}
\ee
where $\mathbf I$ is the $n_\theta\times n_\theta$ identity matrix. Its covariance is therefore
\be
  \mathbf M=
  \mathbf A\,\mathbf C_{\mathbf d}\,\mathbf A^\top.
  \label{eq:ratio_metric}
\ee

The weights that minimize Eq.~\eqref{eq:ratio_var} are proportional to
$\mathbf M^{-1}\boldsymbol{\xi}_{\rm den}^{\rm fid}$\footnote{In practice, we first construct the residual covariance over the full angular range, $\mathbf M_{\rm full}$, and restrict its inverse to the selected bins:
$\mathbf v\propto(\mathbf M_{\rm full}^{-1})_{SS}
\boldsymbol{\xi}_{\rm den}^{\rm fid}$, where $S$ denotes the selected angular bins and $\boldsymbol{\xi}_{\rm den}^{\rm fid}$ is restricted to that range. The implemented weights increase the standard deviation by at most $0.24\%$ relative to the optimal choice for the retained LSST Y10 ratios.}. Using fiducial rather than measured quantities in the weights keeps both sums in Eq.~\eqref{eq:ratio_estimator} linear in the data; weighting with the measured denominator would make the denominator quadratic in the noise and bias the ratio toward zero.

We test the dependence on the assumed fiducial cosmology by assuming an alternative cosmology and find it is negligible: generating the data vector and covariance at $w_0=-0.83$, $w_a=-0.59$ and analyzing them with the weights from the fiducial $\Lambda$CDM cosmology shifts the point estimates by at most $0.003\sigma_{\hat r}$.

To leading order, in the scale-independent  limit, the estimator variance  is
\be
  \sigma_{\hat r}^{2}\simeq
  \frac{\mathbf v^\top\mathbf M\,\mathbf v}
       {(\mathbf v^\top
       \boldsymbol{\xi}_{\rm den}^{\rm fid})^2}.
  \label{eq:ratio_var}
\ee

Note that previous lensing-ratio analyses compressed their ratios differently: \citet{y3-ratios} and \citet{y6-ggl} used a divide-then-average estimator, the GLS fit of a constant to the per-angular-bin ratios $\xi_{\rm num}(\theta_k)/\xi_{\rm den}(\theta_k)$, while our earlier CMB-lensing-ratio analysis \citep{PratBaxter2019} jointly fit a shared angular shape parameter and per-source-plane amplitudes to the full set of correlation functions via MCMC. Some of our ratios, particularly those with small lens--source redshift separation, have lower signal-to-noise than in those analyses, which motivated the estimator described above. Our new estimator divides only after combining the angular bins, avoiding division by individual noisy denominators.  
Tests on noisy simulated data vectors in Appendix~\ref{app:estimator_noise} show negligible bias and approximately Gaussian sampling distributions for these configurations. That appendix also compares the estimator with that of \citet{y3-ratios}.

\subsection{Lensing ratios covariance}\label{sec:lr_cov}

We compute the covariance of the compressed ratios by
linear propagation of the joint analytic covariance
$\mathbf C_{\boldsymbol{\xi}}$ of the data vector $\boldsymbol{\xi}$ that concatenates all correlation functions
entering the analysis (Sec.~\ref{sec:covariance}):
\be
  \left(\mathbf C_{\mathbf r}\right)_{ab}
  = \mathbf g_a^\top\mathbf C_{\boldsymbol{\xi}}\mathbf g_b,
  \label{eq:ratio_cov_lin}
\ee
where $a,b$ index the ratio combinations and  $\mathbf g_a$ are the response vectors, defined by their action on a small shift $\delta\boldsymbol{\xi}$ of the data vector:
\be
  \mathbf g_a^\top\delta\boldsymbol{\xi}
    = \frac{\mathbf v_a^\top\left(\delta\boldsymbol{\xi}_{{\rm num},a}
      - \mathbf R_a\,\delta\boldsymbol{\xi}_{{\rm den},a}\right)}{S_a},
  \qquad
  S_a = \mathbf v_a^\top
      \boldsymbol{\xi}^{\rm fid}_{{\rm den},a},
\ee
so $\mathbf g_a$ is non-zero only in the rows of $\boldsymbol{\xi}$ that hold the numerator and denominator of combination $a$. 
These response vectors approximate the gradient of
Eq.~\eqref{eq:ratio_estimator}. Its exact gradient would 
use the scalar compressed fiducial ratio in place of
$\mathbf R_a$. The two expressions coincide in the
scale-independent limit; for the retained LSST Y10 ratios,
using the exact gradient changes the standard deviations
by at most $0.08\%$. The diagonal of $\mathbf C_{\mathbf r}$ equals
Eq.~\eqref{eq:ratio_var} for the adopted fiducial weights.

\begin{figure*}
    \centering
    \includegraphics[width=\textwidth]{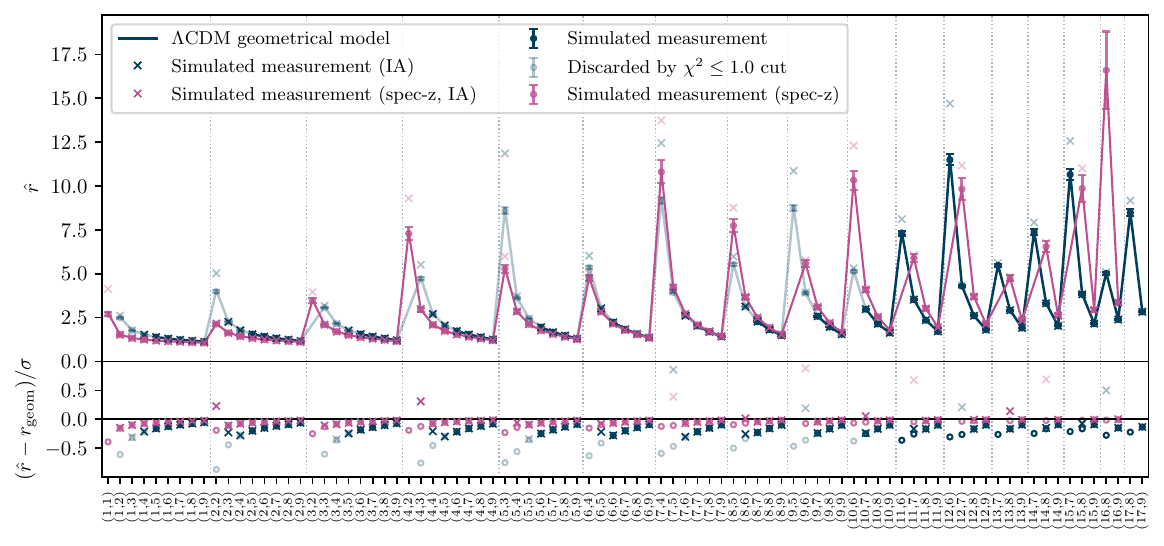}
    \caption{\textbf{Lensing-ratio data vector:}  for the LSST~Y10 photometric lens sample (blue) and the spectroscopic lens sample of Appendix~\ref{app:lens_specz} (pink), all paired with LSST~Y10 sources. Points show the measured ratios, grouped by lens bin $i$ (dotted vertical lines) and ordered by increasing source bin $j$; lines show the geometry-only prediction $r_{\rm geom}$.  Non-geometric combinations removed by the iterative $\chi^2 \leq 1$ selection cut are shown at reduced opacity (the spectroscopic selection retains all its combinations by construction). Crosses show the measurements corresponding to the  IA-contaminated data vectors. The bottom panel shows the residual. }
    \label{fig:datavector_basis1}
\end{figure*}

\subsection{Selecting the geometrical ratios}\label{sec:sim_ratios_cov}

We adopt a nonredundant set of CMB lensing ratios
$\mathcal{R}^{\rm CMB}_{(i,j)}$ for all retained lens--source
pairs, accounting for their full covariance. At the level of
the underlying lensing signals, galaxy--galaxy lensing ratios
can be expressed as
\begin{equation}
\mathcal{R}^{\rm GGL}_{(i,j,j')} =
\frac{\mathcal{R}^{\rm CMB}_{(i,j')}}
     {\mathcal{R}^{\rm CMB}_{(i,j)}} ,
\end{equation}
and are therefore not included as additional observables.
We require a minimum mean-redshift separation
$\bar{z}_s-\bar{z}_l \geq 0.05$ to exclude nearly overlapping
lens--source pairs with weak galaxy--galaxy lensing signals
and poorly determined ratio denominators.

Figure~\ref{fig:datavector_basis1} summarizes the simulated data vector. The data vector $\hat{\mathbf{r}}$ is noiseless: each ratio is computed from the full cross-correlation model of Sec.~\ref{sec:xcorr} (the \textsc{CosmoSIS} integral of Eq.~\eqref{eq:cl}, including galaxy bias, the nonlinear matter power spectrum, and the full lensing kernels). We compare this full-model data vector against the geometry-only prediction $r_{\rm geom}$, computed as in Sec.~\ref{sec:geometry_grid} directly from the critical-surface-density factors of Eqs.~\eqref{eq:ratio_ggl}--\eqref{eq:ratio_cmb}. The figure includes our fiducial LSST Y10 lens sample and an alternative spectroscopic sample we consider in Appendix~\ref{app:lens_specz}.

We apply an iterative $\chi^2$ selection cut to remove ratio combinations where the geometry-only prediction is not valid. Starting from the full set, we compute $\chi^2 = \Delta\mathbf{r}^\top \hat{\mathbf{C}}_{\mathbf{r}}^{-1} \Delta\mathbf{r}$, where $\Delta\mathbf{r} = \hat{\mathbf{r}} - \mathbf{r}_{\rm geom}$, identify the combination contributing the largest $\Delta\chi^2$ upon removal, remove it, and iterate until $\chi^2 \leq \chi^2_{\rm thr}$. We adopt $\chi^2_{\rm thr} = 1$, which retains 59 of the 84 candidate LSST~Y10 ratio combinations (and all 79 spectroscopic-sample combinations: the finer low-redshift lens binning of that sample, described in Appendix~\ref{app:lens_specz}, was chosen precisely so that no combination is discarded). 

Note that the discarded combinations are not unusable: they could be retained by modeling the full cross-correlations with their complete redshift kernels, as in the DES shear-ratio analyses \citep{y3-ratios}, in which galaxy bias and matter clustering still partially cancel. In this work we discard them instead to preserve the purely geometric interpretation of the data vector.

The retained LSST~Y10 ratios reach per-ratio precisions of $\sigma_r/|r| = 0.8$--$2.8\%$ (median $1.0\%$) and a total signal-to-noise of $360$ ($166$ for the spectroscopic sample); Appendix~\ref{sec:error_budget} decomposes the noise budget. Most of this signal-to-noise comes from the small scales that a $3\times2$pt analysis discards: restricting the compression to $\theta$ above $6\,h^{-1}$Mpc comoving ($7$--$31$ arcmin across the lens bins, the DES Y6 galaxy--galaxy lensing cut) would lower the LSST~Y10 total to $98$.  For the LSST~Y1 configuration the same selection retains 31 of 37 valid combinations, with per-ratio precisions of $1.5$--$4.5\%$ (median $2.0\%$) and a total signal-to-noise of $163$.

\subsection{External simulated likelihoods}\label{sec:external_likelihoods}

To quantify how much the lensing ratios would add to current geometric probes, we will combine them with different combinations of external simulated data, which we describe in this section. All external data vectors are simulated at the fiducial cosmology (Table~\ref{tab:fiducial}) to ensure a self-consistent forecast.

\paragraph*{DES Year 5 Type Ia supernovae.}
We use a simulated version of the DES Year 5 supernova Hubble diagram \citep{DESCollab2024SN}, with 1829 SNe over 0.025 < z < 1.12. The likelihood is a $\chi^2$ on the distance moduli $\mu(z_i)$, analytically marginalized over the supernova absolute magnitude, with the statistical and systematic covariance of the real DES Y5 release. Supernovae constrain the luminosity distance as a function of redshift, $D_L(z)=(1+z)f_K[\chi(z)]$, providing sensitivity to the dark energy equation of state through the late-time expansion rate.

\paragraph*{DESI DR2 BAO.}
  We use a simulated version of the DESI Data Release 2 baryon acoustic oscillation measurements \citep{DESIY2BAO2025}.
  The DESI DR2 data set comprises 13 distance measurements spanning $0.3 \lesssim z_\mathrm{eff} \lesssim 2.3$ and providing ratios of the comoving angular diameter distance $D_M$ and the Hubble distance $D_H \equiv c/H(z)$ to the sound horizon at the baryon drag epoch, $r_d$.  BAO measurements are sensitive to the integrated expansion history and constrain $\Omega_m$ and $h\,r_d$.

  \paragraph*{Angular scale of the sound horizon ($\theta_\star$ prior).}
 A Gaussian prior on the angular scale of the sound horizon at recombination, $\theta_\star = r_\star / D_M(z_\star)$, where $r_\star$ is the sound horizon at photon decoupling, $z_\star$. When we use this likelihood, we  compute $\theta_\star$ with CAMB \citep{Lewis2000CAMB}, run in a mode that solves only the background expansion and the recombination history, without perturbations: $z_\star$ and $r_\star$ come from the recombination calculation, and $D_M(z_\star)$ from the background, including  the massive-neutrino contribution. The same recombination calculation provides the BAO sound horizon $r_d$, evaluated at the drag epoch instead of $z_\star$. The prior mean is set to the value of this calculation at the fiducial cosmology, $\theta_\star^{\rm fid} =
  0.0104081$. Following DESI DR2, we adopt the width $\sigma = 5.3\times10^{-6}$, which is 75\% broader than the nominal \textit{Planck} 2018 precision to account for potential model dependence in extended cosmologies \citep{DESIY2BAO2025}.

\paragraph*{Compressed CMB prior.}
Compressed CMB prior of \citet{LemosLewis2023}, used in the DESI DR2 analysis \citep{DESIY2BAO2025}, which is a joint Gaussian on $(\theta_\star,\;\omega_b,\;\omega_{bc})$. Here $\omega_{bc} = (\Omega_b + \Omega_c)\,h^2 = (\Omega_m - \Omega_\nu)\,h^2$ is the physical cold-matter (baryon + CDM) density, excluding the neutrino contribution. The prior means are set to the fiducial values $\boldsymbol{\mu}^{\rm fid} = (0.0103778,\;0.022264,\;0.142172)$, computed at the fiducial cosmology  accounting for the neutrino mass in both the sound-horizon integral and the angular diameter distance; $\theta_\star$ here is evaluated with the same analytic sound-horizon computation used by the likelihood, and therefore differs slightly from the Boltzmann-code value quoted above, and the covariance is taken from \citet{LemosLewis2023}. We adopt this compressed prior rather than a full CMB likelihood because it requires no perturbation computation: like the lensing ratios, BAO and SNe, it depends only on the background expansion and the recombination history, so the full analysis remains inexpensive. This compressed prior keeps most of the constraining power of the full CMB likelihood: combined with BAO and SNe, it gives $\mathrm{FoM}(w_0, w_a) = 189$ (Table~\ref{tab:lr_fom}), compared with $202$ for the full likelihood \citep{DESY6w0wa2026}.

\section{Results}\label{sec:results}

We consider five extended late-time model spaces ($\Lambda$CDM~+~$\Omega_k$, $w$CDM~+~$\Omega_k$, flat $w_0w_a$CDM, $w_0w_a$CDM~+~$\Omega_k$, and $w_0w_a$CDM~+~$\Omega_k$~+~$\sum m_\nu$), and the non-parametric binned dark-energy reconstruction of Sec.~\ref{sec:results_binned_de}. All data vectors are simulated at the flat-$\Lambda$CDM fiducial of Table~\ref{tab:fiducial}. All chains use the \textsc{nautilus} sampler ~\citep{Lange2023} and because the pipeline only requires background quantities, they are inexpensive (on 256 CPU cores a typical chain  costs one to two node-hours, and the most expensive chains, which add the 35 photo-$z$ and shear-calibration nuisance parameters of Appendix~\ref{app:systematics}, five to ten).

We quote figures of merit $\mathrm{FoM}(p_1, p_2) = 1/\sqrt{\det \mathrm{Cov}}$ of the marginalized 2D posteriors. For each model we evaluate the FoM in a single reference plane, spanned by the parameters that extend the model beyond flat $\Lambda$CDM: $(w_0, w_a)$ for the $w_0w_a$ models, $(w, \Omega_k)$ for $w$CDM~+~$\Omega_k$, and, for $\Lambda$CDM~+~$\Omega_k$, whose only extension parameter is $\Omega_k$, the $(\Omega_m, \Omega_k)$ plane. We refer to these as the headline planes below. 

\begin{figure*}
    \centering
    \includegraphics[width=\textwidth]{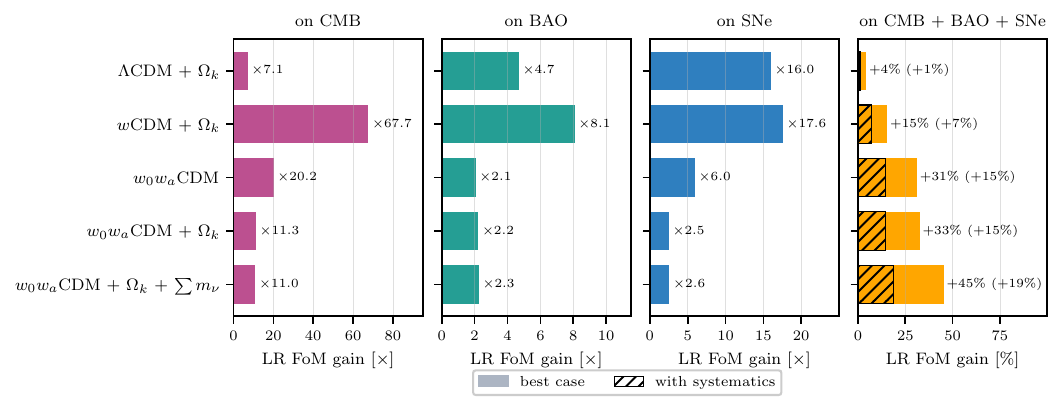}
    \caption{\textbf{The lensing-ratio gain across models.} Gain in the figure of merit (FoM) in each model's headline plane (see Tab.~\ref{tab:lr_fom}) from adding the lensing ratios to each external data set. The first three panels give the factor by which the ratios raise the FoM of the compressed CMB prior, the DESI DR2 BAO and the DES Y5 SNe on their own in the best-case scenario. The right panel gives the fractional gain on the full CMB + BAO + SNe combination, with filled bars for the best-case configuration  and hatched bars after marginalizing over systematics.}
    \label{fig:gain_summary}
\end{figure*}

\begin{table*}
\centering
\caption{Figures of merit ${\rm FoM} = 1/\sqrt{\det \mathrm{Cov}}$ in each model's headline parameter plane, for the lensing ratios (LR)
alone and in combination with external data. CMB denotes the compressed CMB prior of \citet{LemosLewis2023}, a joint Gaussian on $(\theta_\star, \omega_b, \omega_{bc})$ (Sec.~\ref{sec:external_likelihoods}); BAO and SNe denote the simulated DESI DR2 and DES Y5 likelihoods; ``All'' is CMB+BAO+SNe+LR. The lensing ratios are in their best-case configuration with systematic-error parameters fixed and the optimized lens and source samples (marginalized cases in Table~\ref{tab:lr_systematics}). The last column repeats the full combination under the final systematics treatment of Appendix~\ref{app:systematics} (IA-contaminated data vector, photo-$z$ and shear calibration marginalized).}
\label{tab:lr_fom}
\begin{tabular}{llrrrrrrrrrr}
\hline
model & plane & \multicolumn{1}{c}{LR} & \multicolumn{1}{c}{CMB} & \multicolumn{1}{c}{CMB+LR} & \multicolumn{1}{c}{BAO} & \multicolumn{1}{c}{BAO+LR} & \multicolumn{1}{c}{SNe} & \multicolumn{1}{c}{SNe+LR} & \multicolumn{1}{c}{CMB+BAO+SNe} & \multicolumn{1}{c}{All} & \multicolumn{1}{c}{All (with sys.)} \\
\hline
    $\Lambda$CDM + $\Omega_k$ & $\Omega_m$--$\Omega_k$ & 938.6 & 7\,250 & 51\,664 & 2\,883 & 13\,523 & 530.9 & 8\,489 & 221\,371 & 230\,929 & 223\,845 \\
    $w$CDM + $\Omega_k$ & $w$--$\Omega_k$ & 39.3 & 63.7 & 4\,307 & 233.4 & 1\,886 & 38.3 & 673.8 & 34\,219 & 39\,512 & 36\,631 \\
    $w_0w_a$CDM & $w_0$--$w_a$ & 8.7 & 2.9 & 59.0 & 11.7 & 24.3 & 5.8 & 34.4 & 189.3 & 248.4 & 216.9 \\
    $w_0w_a$CDM + $\Omega_k$ & $w_0$--$w_a$ & 1.2 & 1.0 & 10.9 & 8.5 & 19.0 & 3.6 & 8.9 & 166.4 & 220.7 & 190.9 \\
    $w_0w_a$CDM + $\Omega_k$ + $\sum m_\nu$ & $w_0$--$w_a$ & 1.2 & 1.0 & 10.6 & 8.6 & 19.5 & 3.5 & 9.0 & 111.1 & 161.4 & 131.8 \\
\hline
\end{tabular}
\end{table*}

\subsection{Impact of the lensing ratios across extended dark-energy and curvature models}\label{sec:results_baseline}

We first present the results that correspond to the best-case configuration of the forecast: the optimized LSST Y10 lens and source samples, with all systematic-error parameters fixed at their input values. Figure~\ref{fig:gain_summary} and Table~\ref{tab:lr_fom} summarize the gains lensing ratios add across the models. We obtain the following results:

\begin{itemize}
  \item The ratios alone constrain  ratios of distances, so their standalone figures of merit are modest in absolute terms, though competitive with the other single probes. 
  \item Combined with the CMB prior they become a competitive dark-energy and curvature probe: its physical densities $(\omega_b, \omega_{bc})$ calibrate $\Omega_m h^2$ and convert the ratios' distance-ratio information into absolute parameter constraints, improving the CMB-only figures of merit by factors of $7$--$68$ depending on the model. We find they are also very complementary to BAO and SNe alone. 
  \item On top of the full combination of CMB + BAO + SNe the ratios contribute significantly: $+33\%$ (open $w_0w_a$), $+31\%$ (flat $w_0w_a$) and $+15\%$ ($w$CDM~+~$\Omega_k$) in the dark-energy planes ($+4\%$ in the $\Lambda$CDM~+~$\Omega_k$ curvature plane). With the neutrino mass additionally freed, the open-$w_0w_a$ contribution rises to $+45\%$ ($+36\%$ with free neutrino mass but fixed curvature)\footnote{Note that the gain is due to degeneracy breaking rather than sensitivity to the neutrino mass itself. We find that adding the ratios reduces the correlation between $\sum m_\nu$ and $w_a$ from $-0.40$ to $-0.28$ and $\sigma(w_a)$ shrinks from $0.29$ to $0.23$ (with fixed curvature from $-0.43$ to $-0.37$ and from $0.25$ to $0.21$).}. Thus, overall we find that the freer the model space, the more LRs contribute to the full combination, highlighting the value of adding more probes in extended model spaces. 
\end{itemize}

All the gains quoted above assume the lensing-ratio systematic-error parameters fixed at their input values. Switching on the three leading measurement systematics as described in Appendix~\ref{app:systematics} (one mean-redshift shift per tomographic bin under the priors of Table~\ref{tab:cases_updated}, one multiplicative shear bias per source bin, and an unmodelled intrinsic-alignment signal injected into the data vector at the DES~Y6 3x2pt best-fit TATT amplitude) costs the ratios roughly half of their standalone constraining power in flat $w_0w_a$CDM ($\mathrm{FoM}(w_0, w_a) = 8.7 \to 4.1$). The cost comes from the shear and photometric-redshift calibration priors rather than by the intrinsic alignment contamination, and we find that the lens mean-redshift shifts  self-calibrate, tightening to $0.36$--$0.79$ of their prior width. We find that when marginalizing over forecasted systematics, their gain on the  CMB + BAO + SNe combination retains about half of the best-case gain in every model, as shown in the right panel of Figure~\ref{fig:gain_summary} in hatched bars.

Finally, we also check whether the gains depend on the $\Lambda$CDM input cosmology: with every probe re-simulated at the DES~Y6 all-data $w_0w_a$CDM highest-posterior sample  \citep{DESY6w0wa2026} ($w_0 = -0.83$, $w_a = -0.59$), we find the ratios add $+34\%$ to the CMB + SNe + BAO combination in flat $w_0w_a$CDM, vs $+31\%$ at the fiducial input $\Lambda$CDM.

\begin{figure*}
    \centering
    \includegraphics[width=\textwidth]{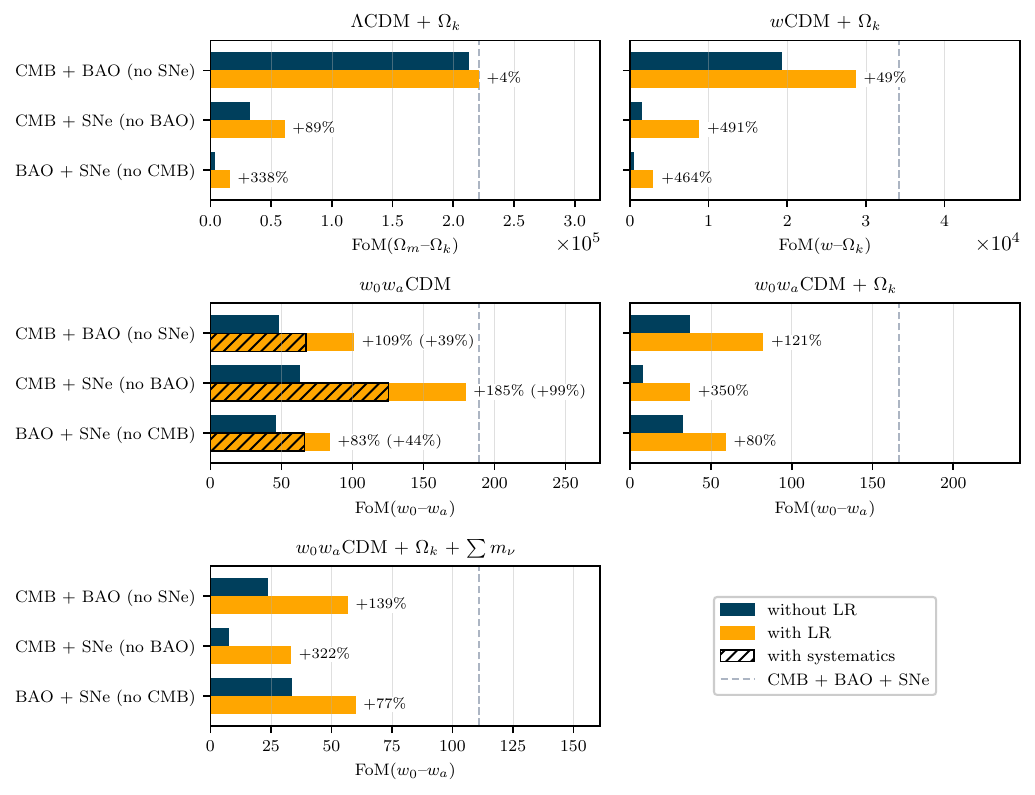}
    \caption{\textbf{Leave-one-out combinations.} Figures of merit in each model's headline plane for pairs that drop one of the three external probes from the full CMB + BAO + SNe combination. The dashed line marks the  CMB + BAO + SNe combination without the ratios, so each yellow bar shows how much of the information lost with the dropped probe the ratios put back; the labels give the fractional FoM gain from adding them. Hatched bars repeat the flat $w_0w_a$CDM pairs marginalizing over systematics, with the gain in parentheses. }
    \label{fig:lr_loo}
\end{figure*}

\begin{figure*}
    \centering
    \includegraphics[width=\textwidth]{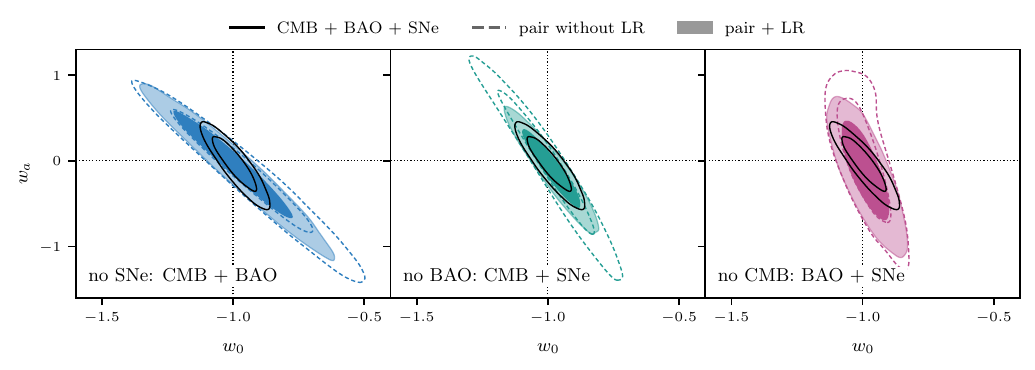}
    \caption{\textbf{Leave-one-out combinations in the flat $w_0w_a$CDM model.} Each panel drops one probe from the full combination of CMB + BAO + SNe (shown in black). Dashed contours show the remaining pair of external probes on its own, and filled contours show it after adding the lensing ratios under the final systematics treatment of Appendix~\ref{app:systematics} (IA-contaminated data vector, photo-$z$ and shear calibration marginalized). Adding lensing ratios raises the CMB + SNe
figure of merit from 33\% to 66\% of the full
CMB + BAO + SNe baseline, enabling a more competitive
leave-one-out robustness test. }
    \label{fig:lr_loo_contours}
\end{figure*}

\subsection{Independent constraints on the expansion history}\label{sec:results_geometry}

In this section we test the power of lensing ratios to yield independent constraints on the expansion history, by providing an additional geometrical probe that enables competitive leave-one-out (dropping one external probe at a time from CMB + BAO + SNe) robustness tests.  We present the results across the different models in Fig.~\ref{fig:lr_loo}. 

We find that when removing one probe at a time, the  remaining pairs  maintain  $2$--$57\%$ of the full CMB + BAO + SNe combination (with only  one exception for  CMB + BAO in $\Lambda$CDM + $\Omega_k$ which reaches $96\%$). We find that adding the ratios is able to regain a significant part of the constraining power lost in most cases. Two specific cases are especially interesting: 1) in flat $w_0w_a$CDM, where the ratios added to the CMB + SNe reach $95\%$ of the full FoM (instead of 33\% without the ratios), nearly substituting for the BAO and 2) CMB + BAO in $w$CDM + $\Omega_k$, where the ratios recover significant FoM for the missing SNe.

We also show the three flat $w_0w_a$CDM pairs under the final systematics treatment (hatched bars in the $w_0w_a$CDM panel). The gains roughly halve, in line with what we find in Figure~\ref{fig:gain_summary}. Figure~\ref{fig:lr_loo_contours} shows these three combinations after marginalizing for the systematics in the $w_0$--$w_a$ plane, together with the same pairs without the ratios.

\subsection{Non-parametric reconstruction of the dark-energy evolution}\label{sec:results_binned_de}\label{sec:binned_de}

\begin{figure*}
    \centering
    \includegraphics[width=\textwidth]{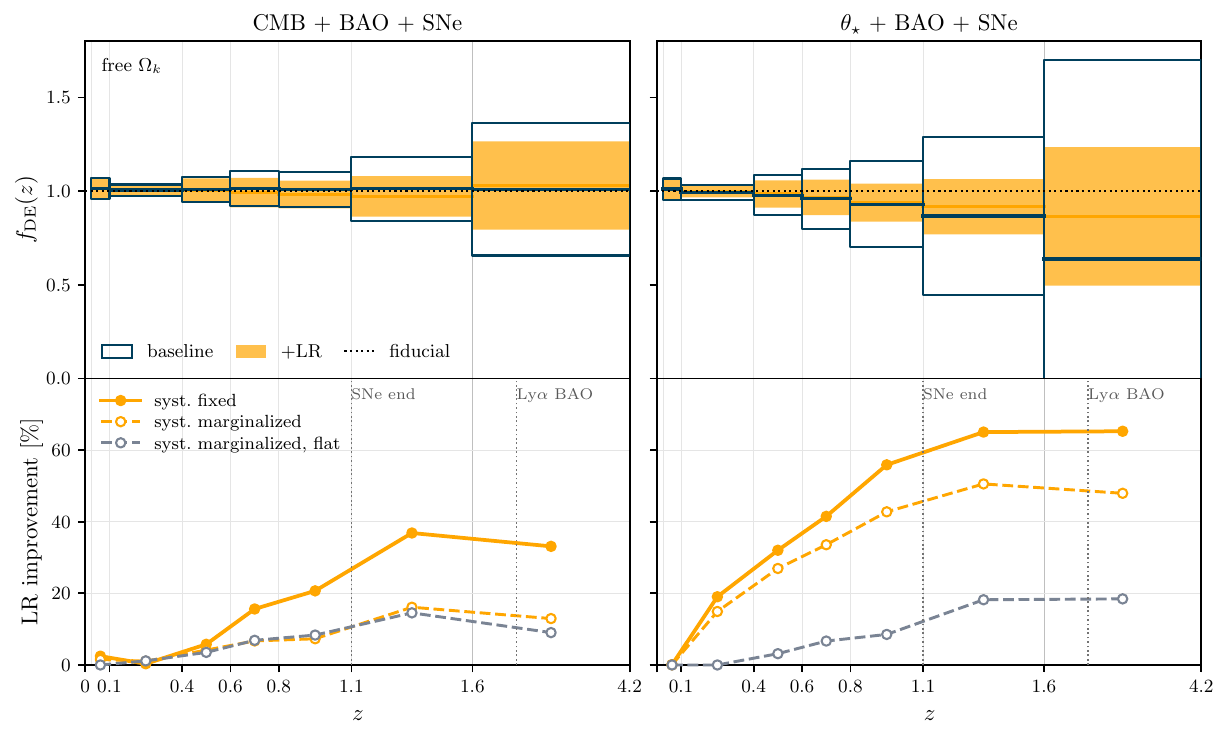}
    \caption{\textbf{Non-parametric reconstruction of the dark-energy density and the lensing-ratio contribution.} Top: $f_{\rm DE}(z) = \rho_{\rm DE}(z)/\rho_{\rm DE}(0)$ in seven redshift bins with free curvature, from simulated data vectors at the $\Lambda$CDM fiducial. Lines show the posterior mean per bin and bands the 68\% intervals: unfilled for the baseline combination, filled after adding the LSST~Y10 lensing ratios (+LR), for the compressed CMB prior (left; $\theta_\star$, $\omega_b$, $\omega_{bc}$) and the $\theta_\star$-only anchor (right). Bottom: the per-bin improvement from the ratios, with the systematic-error parameters fixed (filled markers), marginalizing over them (open markers; Appendix~\ref{app:systematics}). The grey dashed line shows the marginalized case with flat geometry. Dotted vertical lines mark the reach of the SNe Hubble diagram ($z \simeq 1.1$) and the last BAO point from Lyman-$\alpha$   ($z_{\rm eff} = 2.33$). The $z$ axis is linear up to $z = 1.6$ and compressed by a factor of four across the last bin. }
    \label{fig:fde_bins}
\end{figure*}

The evidence for dynamical dark energy is framed almost entirely within the $w_0w_a$ parametrization, which imposes a time dependence that does not correspond directly to any physically motivated model. Instead, non-parametric reconstructions are able to constrain the evolution with minimal assumptions about its time dependence, at the cost of a weaker signal per parameter, which is precisely where additional geometric information is most valuable.

Following \citet{Kessler2026}, we reconstruct the dark-energy density in redshift bins, writing
\be
\begin{split}
  H^2(z) = H_0^2\big[&\Omega_m(1+z)^3 + \Omega_r(1+z)^4 \\
  &+ \Omega_k(1+z)^2 + \Omega_{\rm DE}\,f_{\rm DE}(z)\big],
\end{split}
  \label{eq:Hz_binned}
\ee
where $f_{\rm DE}(z)\equiv\rho_{\rm DE}(z)/\rho_{\rm DE}(0)$ is the dark-energy density relative to its present value, related to the equation of state $w(z)$ by
\be
  f_{\rm DE}(z) = \exp\!\left[3\int_0^z \frac{1+w(z')}{1+z'}\,dz'\right].
  \label{eq:fde_def}
\ee
Our binned reconstruction treats the density
amplitudes as phenomenological parameters, allowing negative
values, with flat priors in the range $f_{\rm DE}(z) \in [-1,3]$ (therefore it does not need to admit an equivalent regular
$w(z)$ description).

We align the bin edges with the DESI tracer redshift ranges, $\{z_i\}=\{0.025,0.1,0.4,0.6,0.8,1.1,1.6,4.2\}$, and fix $\Lambda$CDM behaviour ($w=-1$, $f_{\rm DE}=1$) beyond $z_{\rm max}=4.2$. We take the density to be piecewise constant, $f_{\rm DE}(z)=f_i$ for $z\in[z_i,z_{i+1})$. Below $z_{\rm min}=0.025$, we fix $f_{\rm DE}=1$,
consistent with its normalization at $z=0$. Extending
this value across the interval is a modeling choice,
motivated by the weak distance sensitivity to a free
amplitude over such a narrow slice. We sample the seven amplitudes jointly with $(\Omega_m, h, \Omega_b)$ and, where noted, $\Omega_k$ and $\sum m_\nu$. We compare this choice with binning the equation of state in Sec.~\ref{sec:fvsw_choice}.

The left and bottom panels of \fig{fig:fde_bins} show the per-bin constraints and the lensing-ratio contribution on the compressed CMB prior with the simulated DESI DR2 BAO and DES Y5 SNe baseline, for the free-curvature model. At $z \lesssim 0.6$, where the BAO and SNe data are most constraining, the ratios add $\lesssim 5\%$; their contribution rises to $16$--$21\%$ at $0.6 < z < 1.1$ and to $33$--$37\%$ in the two highest-redshift bins. The binned parametrization thus localizes in redshift the gains of the previous sections: the ratios contribute where the SNe Hubble diagram ends ($z \simeq 1.1$) and the BAO tracers thin out, reaching the last bin only through the Lyman-$\alpha$ points at $z_{\rm eff} = 2.33$.

The contribution of the ratios grows by a factor of $2.4$ when the CMB information is reduced to its minimal, background-only content, the acoustic scale $\theta_\star$ alone (right and bottom panels of \fig{fig:fde_bins}). $\theta_\star$ is the CMB quantity most robust to cosmological assumptions \citep{PlanckParams2018, LemosLewis2023}, whereas $\omega_b$ and $\omega_{bc}$ are more model dependent; omitting them keeps the reconstruction driven by the late-time distances alone, insensitive to early-time physics and to a possible matter-density tension between early- and late-time probes, which is the reasoning behind the baseline configuration of \citet{Kessler2026}. With free curvature the $\theta_\star$ baseline inflates by $35$--$140\%$ per bin (the last bin reaching $\sigma = 1.06$), and the ratios recover most of this loss: the mean improvement rises to $39.8\%$, reaching ${\approx}65\%$ at $z > 1.1$, and $\sigma(\Omega_k)$ tightens by a factor of two, from $0.0099$ to $0.0046$.

The ratios matter more with the $\theta_\star$-only anchor because, without $\omega_b$ and $\omega_{bc}$, the sound horizon is no longer calibrated. The BAO and SNe then constrain only the relative, low-redshift expansion, and the single integrated distance to recombination fixed by $\theta_\star$ can be held constant while $\Omega_k$, the high-redshift bins and $(\Omega_m, h)$ trade against each other. The CMB lensing ratios are free of any ruler: through $f_K(\chi_\star - \chi_l)/f_K(\chi_\star)$ they measure the distance from each lens redshift to recombination, adding one such integrated constraint per lens bin with a different dependence on curvature, which breaks the degeneracy (Fig.~\ref{fig:fde_ok_degeneracy} shows the last redshift bin).

With curvature fixed, the compressed-CMB reconstruction behaves similarly to the free-curvature case ($15\%$ versus $16\%$ mean improvement), and on the $\theta_\star$-only anchor the ratios' contribution drops to $17\%$ on average ($38$--$39\%$ in the two highest-redshift bins), close to the compressed case.

Finally, we quantify the cost of marginalizing over systematics for the reconstruction, repeating the prescription of Appendix~\ref{app:systematics}, IA-contaminated data with the photo-$z$ and shear-calibration parameters marginalized (see open markers in the bottom panel of \fig{fig:fde_bins}). We find that for the CMB + BAO + SNe baseline the ratios' contribution halves to a $6$--$7\%$ mean improvement, with $9$--$16\%$ in the two highest-redshift bins. On the minimal $\theta_\star$ anchor  the mean improvement is $31\%$ (versus $40\%$ with systematics fixed), still reaching $48$--$51\%$ in the two highest-redshift bins, with $\sigma(\Omega_k) = 0.0055$ versus $0.0046$. 
The  cost is dominated by the shear calibration of the highest source bins, which is degenerate with the last density bin.  Appendix~\ref{app:source_opt} shows that the standard SRD source sample performs almost as well for the reconstruction.

\subsubsection{Dark energy density bins versus equation-of-state bins}
\label{sec:fvsw_choice}

An alternative to density binning is to take $w(z)=w_i$,
which gives a continuous $f_{\rm DE}(z)$ through
Eq.~\eqref{eq:fde_def}. We compare the two parametrizations
using the Fisher setup of Appendix~\ref{app:source_opt},
with additional chain validation for the free-curvature,
$\theta_\star$-only case. Systematic-error parameters are
fixed throughout this comparison.

\begin{figure}
    \centering
    \includegraphics[width=\columnwidth]{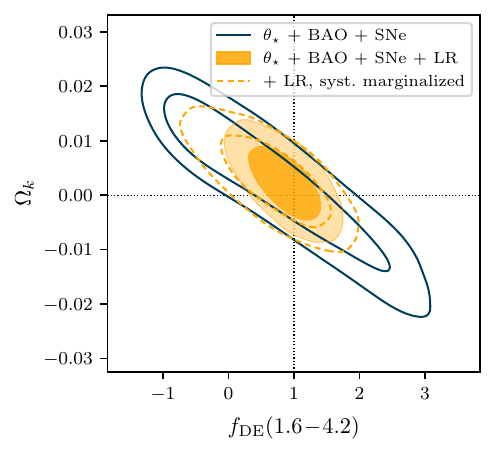}
    \caption{\textbf{Curvature--dark-energy degeneracy.}
    Joint posterior of the highest-redshift density bin and
    spatial curvature for the $\theta_\star$ + BAO + SNe
    reconstruction. Adding lensing ratios (filled contours)
    weakens the degeneracy, changing the correlation
    coefficient from $r=-0.91$ to $-0.64$.
    The dashed contour includes marginalization over
    lensing-ratio systematics
    (Appendix~\ref{app:systematics}).
    Thin dotted lines mark the input cosmology.}
    \label{fig:fde_ok_degeneracy}
\end{figure}

With free curvature, the lensing ratios yield similar gains
in both parametrizations when using the compressed CMB prior.
With only the $\theta_\star$ anchor, however, they reduce
the density-bin uncertainties by about 40\% on average,
compared with 14\% for the equation-of-state bins. In this
configuration, the density bins exhibit stronger curvature
degeneracies, which the ratios help break
(Fig.~\ref{fig:fde_ok_degeneracy}). With curvature fixed,
the average improvements are similar, at 14--16\% for
both parametrizations and either anchor. These comparisons
depend on the adopted parametrizations and priors:
equal-width flat priors in $f_i$ and $w_i$ do not impose
equivalent assumptions on the expansion history.

\subsection{LSST Y1 + SO baseline forecasts}\label{sec:results_y1}

We now ask how much of this program is already available at the LSST Y1 epoch, using the Y1 configuration of Table~\ref{tab:cases_updated}. With systematics fixed, the Y1 ratios reach $\mathrm{FoM}(w_0, w_a) = 2.4$ alone and $17$ with the CMB prior in flat $w_0w_a$CDM ($\sim30\%$ of the Y10 values), and add $+8\%$ to the full CMB + BAO + SNe combination. Marginalizing over the Y1 systematics (IA-contaminated data with photo-$z$ and shear calibration marginalized, using SRD Y1 priors) lowers these to $\mathrm{FoM}(w_0, w_a) = 7.5$ with the CMB prior ($1.5$ in the free-neutrino-mass space) and $+3.5\%$ on the full combination. 

We find that the non-parametric reconstruction of Sec.~\ref{sec:results_binned_de} is where the first-year ratios add  the most. On the minimal $\theta_\star$ anchor, with systematics fixed, they  improve the density bins by $17\%$ on average ($40\%$ at Y10), reaching $23$--$28\%$ at $0.6 < z < 1.6$, and still add $9$--$12\%$ in these bins after marginalizing over the Y1 systematics. The adopted Y1 lens sample stops at $\bar z = 0.94$, so the high-redshift lens--source combinations that carry the Y10 gains are missing. At Y1, the ratios could therefore especially add interesting information only if a lens sample reaching higher redshifts were built.

\section{Conclusions}\label{sec:conclusions}

We have developed and validated an end-to-end lensing-ratio framework for LSST and the Simons Observatory and forecast its cosmological impact in simulated joint analyses with the compressed CMB prior, DESI DR2 BAO and DES Y5 supernovae. In the geometric limit, ratios of galaxy--galaxy lensing and galaxy--CMB lensing around the same lenses cancel the dependence on galaxy bias and the matter power spectrum, leaving only ratios of distances, including distances to the CMB source plane. We characterized where this limit holds and constructed realistic LSST Y10 lens and source samples optimized for the probe (17 narrow lens bins and 9 source bins), paired with Simons Observatory CMB lensing at goal sensitivity. We introduced an average-then-divide estimator with a linearized covariance, and retained only the lens--source combinations with $\bar z_s-\bar z_\ell \geq 0.05$ that pass an iterative $\chi^2 \leq 1$ test of the geometric limit. As alternatives, we assembled a spectroscopic lens sample from 4MOST, DESI and \textit{Euclid}, which removes the lens photometric-redshift uncertainties, and compared our optimized source binning with the standard SRD binning \citep{Mandelbaum2018}.

We combined the simulated ratios with the external data in five late-time models, from $\Lambda$CDM~+~$\Omega_k$ to $w_0w_a$CDM with free curvature and neutrino mass, and in a seven-bin non-parametric reconstruction of the dark-energy density, where we also considered a minimal CMB anchor with $\theta_\star$ only, the geometry-only case. We quantified the impact of the leading measurement systematics by marginalizing over one mean-redshift shift per tomographic bin and one multiplicative shear bias per source bin, after selecting ratios using a data vector contaminated by intrinsic alignments at the DES Y6 best fit. We also repeated the analysis for the LSST Y1 configuration. Our main findings are:

\begin{itemize}

\item \textbf{The sensitivity of LRs to each parameter peaks at different lens redshifts.} They respond most strongly to spatial curvature, followed by matter density, $w$ and $w_a$. The responses to curvature and matter density grow toward high lens redshift, whereas that to $w$ peaks at lower redshift, so ratios spanning a wide lens-redshift range are benefitial (Sec.~\ref{sec:cosmology_sensitivity}).

   \item \textbf{Achievable narrow lens bins keep most ratios geometric at LSST $\times$ SO precision.}
  With 17 narrow lens bins, 59 of the 84 LSST Y10 redshift combinations pass our geometric selection. These ratios can then use small scales (down to $2.5$~arcmin), where we find most of the signal-to-noise lies: $360$ in total, compared with $98$ above the $6\,h^{-1}$Mpc cut used in $3\times2$pt analyses (Sec.~\ref{sec:sim_ratios_cov}).

 \item \textbf{CMB lensing reconstruction noise and cosmic variance, rather than lens density, limit the statistical precision.}
CMB lensing noise dominates at all lens redshifts, followed by cosmic variance and then shape noise; we forecast a median per-ratio uncertainty of $1.0\%$ for LSST Y10 $\times$ SO (Sec.~\ref{sec:sim_ratios_cov}, Appendix~\ref{sec:error_budget}).

  \item \textbf{Our new average-then-divide estimator increases the constraining power of the LRs compared with the DES Y3 estimator, while keeping a Gaussian likelihood.}
We develop a new estimator which combines the angular bins before dividing, so it never divides by a single noisy bin, and its covariance is propagated analytically rather than estimated from simulations. We find that in our configuration, the DES Y3 estimator  inflates its Monte Carlo covariance by a median factor of $1.8$ (up to $96$) when the denominator is noisy. Ours reaches the nominal $68\%$ and $95\%$ coverage. For the LRs alone, this improves the figure of merit by $2$--$14\%$ in the dark-energy planes, and by up to $1.6\times$ for the lower signal-to-noise spectroscopic lens sample (Appendix~\ref{app:estimator_noise}).

  \item \textbf{LRs contribute increasingly to the full CMB + BAO + SNe combination as the model is extended.}
  The best-case figure-of-merit gain grows from $+4\%$ in $\Lambda$CDM~+~$\Omega_k$ to $+45\%$ with evolving dark energy, free curvature and neutrino mass, with $+31\%$ in flat $w_0w_a$CDM. Accounting for systematics typically retains about half of the gain.
  In the seven-bin reconstruction of the dark-energy density, the ratios reduce uncertainties by $\sim35\%$ in the two highest-redshift bins, beyond the reach of the supernovae and where the BAO tracers thin out ($\sim15\%$ after accounting for systematics). In the case with least assumptions, with the geometrical $\theta_\star$ anchor and free curvature, LRs help break the degeneracy between curvature and high-redshift dark energy: the uncertainty reduction reaches  $65\%$ at $z>1.1$ ($50\%$ after accounting for systematics).

  \item \textbf{LRs enable leave-one-out robustness tests of the expansion history.} As an independent geometric probe, they  enable competitive leave-one-out tests, in which one of the probes is dropped, and we find they are most effective at replacing the BAO: adding LRs to CMB+supernova raises the figure of merit from 33\% to 95\% of that of the full CMB+supernova+BAO combination ($66\%$ after marginalizing over systematics).

\item \textbf{Photometric-redshift and shear calibration are the limiting factor, rather than intrinsic alignments}
  They halve the LR-only figure of merit ($8.7 \to 4.1$), while intrinsic alignments cost little. 

  \item \textbf{A spectroscopic lens sample performs almost as well as the photometric LSST Y10 sample}
  After marginalizing over systematics, a spectroscopic 4MOST + DESI + \textit{Euclid} lens sample performs as well as the LSST Y10  one for the LRs alone, but adds less in combination ($+8\%$ versus $+15\%$; Appendices~\ref{app:systematics} and \ref{app:lens_specz}). 

  \item \textbf{LRs need a dedicated lens sample, but not a dedicated source sample.}
  The standard LSST SRD source binning gives the same signal-to-noise and performs nearly as well as our optimized one (Appendix~\ref{app:source_opt}).

  \item \textbf{An LSST first-year analysis already contributes where the external constraints are weakest, while the high-redshift gains require a deeper lens sample.} 
  They reduce the uncertainties in the $\theta_\star$-anchored density bins by 
  $9$--$12\%$ at $0.6<z<1.6$ after accounting for systematics. The adopted Y1 lens sample stops at $\bar z=0.94$, so it lacks the high-redshift lens--source combinations that carry the Y10 gains (Sec.~\ref{sec:results_y1}).

\end{itemize}

Lensing ratios provide a probe of the expansion history that needs no standard candle or ruler and reaches the CMB source plane, at the low computational cost of a likelihood that depends only on background quantities. These gains assume that lens magnification can be suppressed while maintaining the adopted lens densities and redshift distributions, following our procedure on COSMOS (\S\ref{sec:samples}) and the method of \citet{alarcon2026weaklensingbigmagnification}. The contribution from LRs is largest in extended model spaces, beyond the redshift reach of the supernovae, and in leave-one-out tests, where independent geometric information can help test the evidence for evolving dark energy and disentangle it from degeneracies with curvature and neutrino mass.

\section*{Acknowledgments}

JP and CS would like to thank Colin Hill for helpful discussions about SO noise curves, and Alex Alarcon and Gisela Camacho for discussions regarding this topic.

{\small
JP acknowledges support from Project PID2025-172756NA-C33 funded by MICIU/AEI/10.13039/501100011033 and Grant RYC2024-048406-I funded by MICIU/AEI/10.13039/501100011033 and by ESF+. CS acknowledges support from projects PID2022-138123NA-I00 and PID2025-172756NB-C32 funded by MICIU/AEI/10.13039/501100011033 and FEDER, UE, and 
Ramón y Cajal project RYC2021-031194-I, funded by
MCIN/AEI/10.13039/501100011033 and by the “European Union NextGenerationEU/PRTR”. }

\appendix

\section{Estimator validation on noisy data and comparison with the DES Y3 shear-ratio estimator}\label{app:estimator_noise}

In this appendix we validate the ratio estimator of Eq.~\eqref{eq:ratio_estimator} on noisy data, and compare it against the estimator used in the DES~Y3 and Y6 shear-ratio analyses \citep{y3-ratios, y6-ggl},  to connect with previous work and to explain why the present analysis requires a different estimator.

\begin{figure}
    \centering
    \includegraphics[width=\columnwidth]{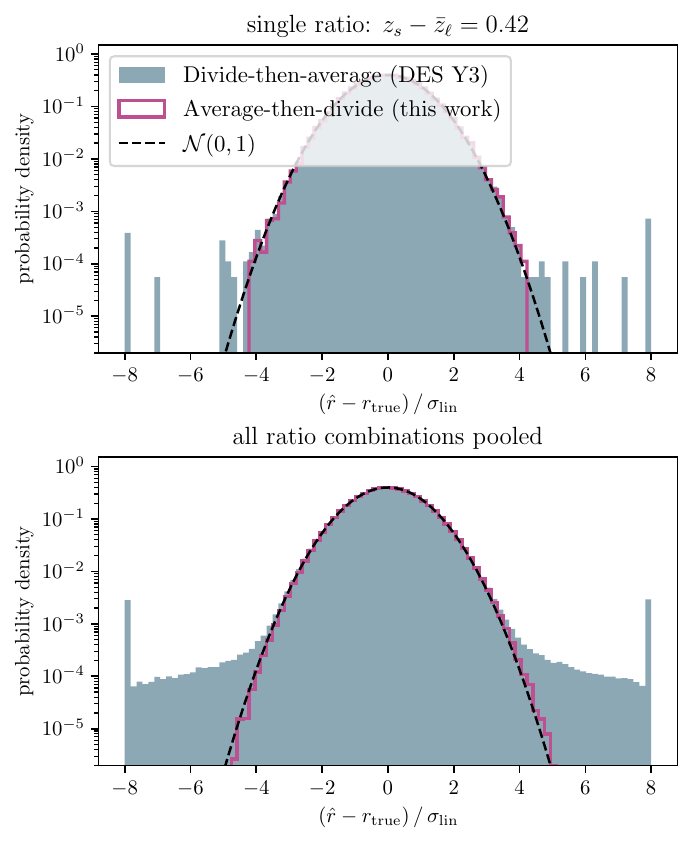}
  \caption{\textbf{Estimator pulls on noisy data.} Pull distributions of the two ratio estimators over $10^5$ noisy simulated data vectors (LSST~Y10), normalized by the linearly propagated uncertainty $\sigma_{\rm lin}$. Top: the lens--source combination with the strongest tails; bottom: all combinations pooled (all 84 ratios, before the $\chi^2$ selection). Our new average-then-divide estimator, Eq.~\eqref{eq:ratio_estimator}, follows the Gaussian (dashed); the divide-then-average estimator of \citet{y3-ratios}, Eq.~\eqref{eq:ratio_estimator_sr}, develops  tails whenever individual denominator bins fluctuate toward zero.}
    \label{fig:estimator_pulls}
\end{figure}

\paragraph{The DES Y3 estimator.}
For a single lens--source combination, both estimators can be written in terms of the same per-angular-bin ratios,
$r_k = \hat\xi_{\rm num}(\theta_k)/\hat\xi_{\rm den}(\theta_k)$, and both are built from weights fixed at
the fiducial model. The estimator of \citet{y3-ratios} averages the ratios directly,
\be
  \hat r_{\rm Y3} = \sum_k w_k\, r_k ,
  \label{eq:ratio_estimator_sr}
\ee
where $w_k$ are the minimum-variance weights for fitting a constant to the $r_k$, given their covariance
propagated from $\mathbf{C}_{\mathbf d}$ (Eq.~\ref{eq:ratio_cov_blocks}). Our estimator,
Eq.~\eqref{eq:ratio_estimator}, also uses fixed weights, $v_k$, but applies them to the two correlation
functions before dividing, rather than to their ratio. It can still be written as an average of the $r_k$:
\be
  \hat r = \sum_k \tilde w_k\, r_k ,
  \qquad
  \tilde w_k \propto v_k\, \hat\xi_{\rm den}(\theta_k) .
  \label{eq:ratio_weights}
\ee
What the two have in common: the fixed weights of \citet{y3-ratios} are $w_k \propto
v_k\,\xi^{\rm fid}_{\rm den}(\theta_k)$,\footnote{To first order, the error on each $r_k$ is the residual of Eq.~\eqref{eq:residual_linear} in that bin
divided by $\xi^{\rm fid}_{\rm den}(\theta_k)$. Averaging the $r_k$ with weights $w_k$ is therefore the same as combining
the residuals with weights $w_k/\xi^{\rm fid}_{\rm den}(\theta_k)$, and the minimum-variance choice for those is $v_k$
(Sec.~\ref{sec:estimator}). Hence $w_k \propto v_k\,\xi^{\rm fid}_{\rm den}(\theta_k)$, provided both estimators use the
same selected angular bins.} so on a noiseless data vector $\tilde w_k = w_k$, and the two
estimators return the same ratio with the same leading-order uncertainty.
Where they differ: first, in their response to noise. In Eq.~\eqref{eq:ratio_estimator_sr}, if the measured
denominator of one angular bin happens to be close to zero, its ratio $r_k$ becomes very large and still enters
with its full weight $w_k$, which produces the heavy tails of Fig.~\ref{fig:estimator_pulls}.
Eq.~\eqref{eq:ratio_estimator} never divides by a single angular bin: the division happens only once, after combining the angular bins. Second, in how the covariance between
ratios is obtained: \citet{y3-ratios} use the sample covariance of the estimator over $10^5$ simulated
realizations, which these heavy tails inflate, whereas we propagate the analytic covariance linearly
(Eq.~\ref{eq:ratio_cov_lin}).

\begin{figure}
    \centering
    \includegraphics[width=\columnwidth]{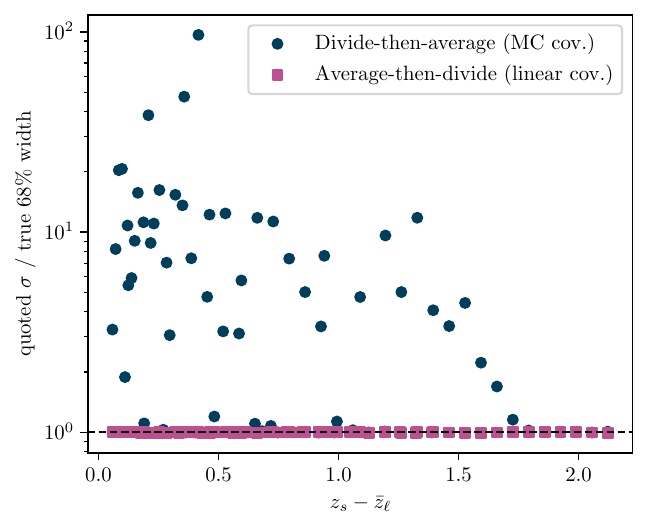}
  \caption{\textbf{Accuracy of the quoted uncertainty.} The ratio of each method's quoted $\sigma$ to the true 68\% width of its sampling distribution as a function of the lens--source separation, for all 84  combinations. The Monte Carlo sample covariance of the divide-then-average estimator is inflated by the tail events by up to two orders of magnitude; the linearly propagated covariance of our estimator is accurate to better than $1\%$ everywhere.}
    \label{fig:estimator_accuracy}
\end{figure}

\begin{figure}
    \centering
    \includegraphics[width=\columnwidth]{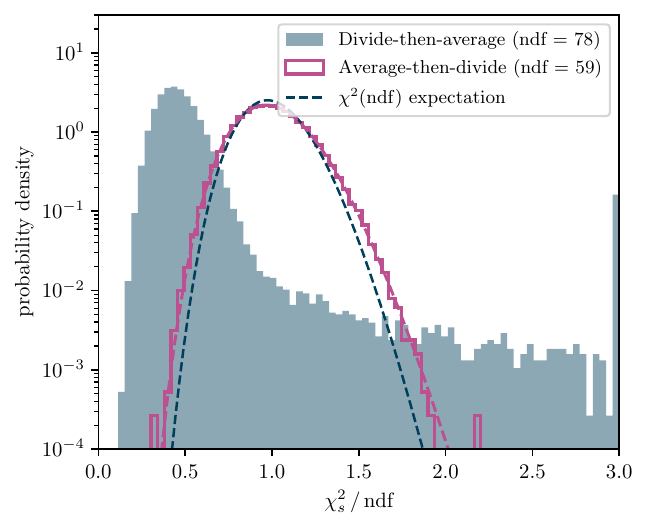}
  \caption{\textbf{Goodness of fit of the two estimators.} Distribution of $\chi^2_s$ between the noisy and true ratios over $10^5$ realizations, normalized by the number of degrees of freedom, for each estimator with its own covariance and its own fiducial $\chi^2 \leq 1$ selection. Dashed and dotted curves show the corresponding $\chi^2({\rm ndf})$ expectations. Our new estimator (magenta) follows the expected distribution; the per-bin-ratio estimator (dark blue) matches its mean by construction but not its shape.}
    \label{fig:estimator_chi2s}
\end{figure}

\paragraph{Validation on noisy data vectors.}
We draw $10^5$ noisy simulated data vectors from the full joint analytic covariance of $[\xi_{g\kappa_g}, \xi_{g\kappa_{\rm CMB}}]$ at the fiducial LSST~Y10 configuration, apply both estimators to every realization and every ratio combination, and compare the recovered ratios against the truth $r_{\rm true}$ computed from the noiseless data vector. The results below use all 84 ratio combinations, before the $\chi^2$ selection. Figure~\ref{fig:estimator_pulls} shows the resulting pull distributions, $(\hat{r} - r_{\rm true})/\sigma_{\rm lin}$, where $\sigma_{\rm lin}$ is the linearly propagated uncertainty from the diagonal of Eq.~\eqref{eq:ratio_cov_lin}, which is to leading order the uncertainty of both estimators (exactly so for Eq.~\eqref{eq:ratio_estimator_sr}, and for Eq.~\eqref{eq:ratio_estimator} in the scale-independent limit). Our new estimator is Gaussian to high accuracy for every combination. The DES-Y3 estimator, by contrast, develops the power-law tails expected for a ratio of Gaussian variables.

The practical consequence is summarized in Fig.~\ref{fig:estimator_accuracy}, which compares each method's quoted uncertainty to the true 68\% width of its sampling distribution. For the Y3 estimator the Monte Carlo sample covariance, dominated by the tail events rather than the width of the distribution, overestimates the true uncertainty by a median factor of $1.8$ and up to a factor of $96$, so the quoted ``$1\sigma$'' interval contains the truth $92.4\%$ of the time (median across combinations). Quoting $\sigma_{\rm lin}$ instead would restore the bulk width but leave the tails unaccounted for, degrading the $2\sigma$ coverage to as low as $92.6\%$. Our new estimator with its linearly propagated covariance has neither problem: its coverage is $68.0$--$68.7\%$ at $1\sigma$ and $95.3$--$95.6\%$ at $2\sigma$ across all combinations, exactly as the Gaussian likelihood assumes.

Figure~\ref{fig:estimator_chi2s} repeats the Gaussianity test that validated the DES~Y3 estimator (fig.~5 of \citealt{y3-ratios}): the distribution of $\chi^2_s = (\mathbf{r}_s - \mathbf{r}_{\rm true})^\top \mathbf{C}_{\mathbf{r}}^{-1} (\mathbf{r}_s - \mathbf{r}_{\rm true})$ over realizations, where $\mathbf{r}_s$ is the ratio data vector of realization $s$ and $\mathbf{C}_{\mathbf{r}}$ is the covariance each estimator would quote (Eq.~\eqref{eq:ratio_cov_lin} for ours, the Monte Carlo sample covariance for the DES~Y3 estimator), computed for each estimator with its own covariance and its own set of combinations retained by the fiducial $\chi^2 \leq 1$ selection cut (59 of the 84 combinations for our estimator and 78 for the DES~Y3 estimator, whose inflated Monte Carlo errors make the cut less restrictive), compared against the expected $\chi^2({\rm ndf})$ distribution. For our new estimator the two agree; for the DES-Y3 estimator the mean matches the number of degrees of freedom by construction (the covariance is the sample covariance of the same realizations) but the distribution does not\footnote{This failure mode is a property of the analysis configuration that is present for lower S/N ratios. In the DES Y3 analyses, 9 ratios were built from the three lowest-redshift lens bins with the highest-redshift source bin as the common denominator, so every denominator bin was measured at high signal-to-noise and the per-angular-bin division never approached zero; the corresponding $\chi^2_s$ test (their fig.~5), equivalent to the one performed here, confirmed a Gaussian likelihood with no noise bias at DES~Y3 precision. Our own test reproduces that conclusion in the same regime: for well-separated lens--source combinations the two estimators are statistically indistinguishable.}. 

\paragraph{Gain in constraining power.}
These sampling properties translate directly into constraining power. We repeated the cosmological inference end-to-end with each estimator, each with its own covariance and its own $\chi^2$ selection (Sec.~\ref{sec:sim_ratios_cov}). These chains  use a  $\chi^2 \leq 0.5$ threshold for both estimators. We compared the resulting figures of merit at matching data combinations. For the ratio-only LSST~Y10 chains the estimator adopted here improves the figure of merit by $2$--$14\%$ in the dark-energy planes and by $22$--$58\%$ in $\Omega_m$--$\Omega_k$ across the four models of Sec.~\ref{sec:results}. The gains are largest for the spectroscopic lens sample, whose lower-signal-to-noise ratios suffer the strongest tail inflation of the Monte Carlo covariance: a factor of $1.6$ in $w_0$--$\Omega_k$ and $2.5$ in $\Omega_m$--$\Omega_k$. Once BAO and SNe are added the difference shrinks to $\lesssim 12\%$. The input cosmology is recovered without significant bias by both estimators.

\section{Lensing ratios with a combined spectroscopic lens sample from 4MOST, DESI and Euclid}\label{app:lens_specz}

\begin{table}
    \renewcommand{\arraystretch}{1.05}
    \centering
    \caption{\textbf{Combined spectroscopic lens sample.} Effective surface number density and $n$-weighted galaxy bias per redshift bin, defined over the single covariance footprint $A=10\,700\,\mathrm{deg}^2$ (4MOST$+$DESI; \textit{Euclid} counted over the $8200\,\mathrm{deg}^2$ it covers within it). These are the inputs to the covariance of Sec.~\ref{sec:covariance}. }
    \label{tab:lens_sample}
    \begin{tabular}{ccc}
        \toprule
        $z$ range & $n_{\mathrm{eff}}$ [arcmin$^{-2}$] & $b_{\mathrm{eff}}$ \\
        \midrule
        $0.100$--$0.175$ & 0.028 & 1.34 \\
        $0.175$--$0.250$ & 0.041 & 1.34 \\
        $0.250$--$0.325$ & 0.040 & 1.35 \\
        $0.325$--$0.400$ & 0.024 & 1.38 \\
        $0.4$--$0.5$ & 0.012 & 1.84 \\
        $0.5$--$0.6$ & 0.014 & 1.93 \\
        $0.6$--$0.7$ & 0.017 & 2.08 \\
        $0.7$--$0.8$ & 0.019 & 2.08 \\
        $0.8$--$0.9$ & 0.033 & 1.92 \\
        $0.9$--$1.0$ & 0.070 & 1.58 \\
        $1.0$--$1.1$ & 0.057 & 1.49 \\
        $1.1$--$1.2$ & 0.054 & 1.53 \\
        $1.2$--$1.3$ & 0.047 & 1.52 \\
        $1.3$--$1.4$ & 0.046 & 1.62 \\
        $1.4$--$1.5$ & 0.040 & 1.61 \\
        $1.5$--$1.6$ & 0.031 & 1.75 \\
        \midrule
        Total & 0.57 & 1.59 \\
        \bottomrule
    \end{tabular}
\end{table}

\begin{figure}
    \centering
    \includegraphics[width=\columnwidth]{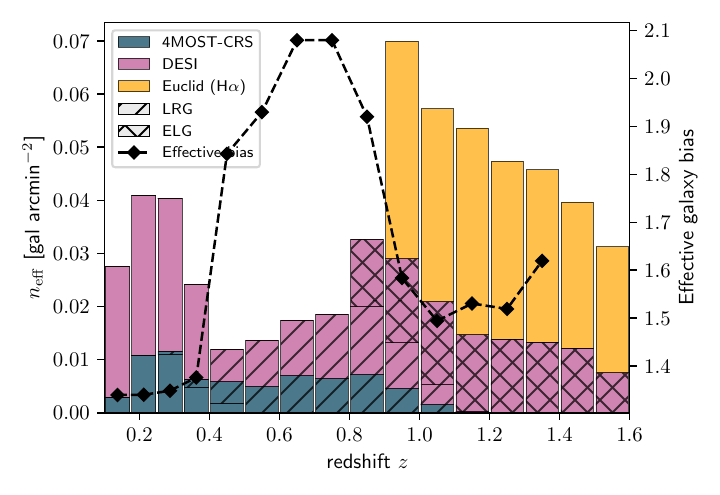}
  \caption{\textbf{Composition of the spectroscopic lens sample.} Decomposition into the contribution of each survey to the effective number density over the $10\,700\,\mathrm{deg}^2$ covariance footprint (bars, left axis), together with the $n$-weighted effective galaxy bias (dashed, right axis). }
    \label{fig:lens_sample}
\end{figure}

We also assemble a spectroscopic lens sample from the three wide spectroscopic surveys that will overlap with LSST to test whether that would be advantageous with respect to the LSST photometric lens sample. For that we consider the 4MOST Cosmology Redshift Survey (CRS) in the south ($A=5700\,\mathrm{deg}^2$), DESI along the equatorial strip ($A=5000\,\mathrm{deg}^2$), and the \textit{Euclid} wide survey, which covers the 4MOST footprint and roughly half of the equatorial strip \citep{Lochner2022} (${\simeq}8200\,\mathrm{deg}^2$ of LSST overlap). Combined that adds to  $A=10\,700\,\mathrm{deg}^2$. We split the sample in redshift bins as displayed in Table~\ref{tab:lens_sample}.

For 4MOST-CRS we digitize the bright-galaxy (BG) and luminous-red-galaxy (LRG) target distributions from fig.~7 of \citet{Verdier2025} 
and multiply by the expected five-year survey completeness of 87\% (BG) and 67\% (LRG) from the survey simulations of the same work.

For DESI we use the DR2 good-redshift counts\footnote{For the LRGs and ELGs we assume the redshift distributions from fig.~4 of \citet{DESIY2BAO2025}. For the BGS galaxies we adopt the full bright-time sample ($>9.1\times10^{6}$ redshifts at $0.1<z<0.4$) rather than the luminosity-cut BAO subsample, with the redshift distribution from the observed $\mathrm{d}N/\mathrm{d}z$ of \citet[fig.~18]{Hahn2023}.} and footprint areas from \citet[tables~2--3 of][]{DESIY2BAO2025}, scaled to the completed five-year survey through the ratio of the expected completed-survey fiber-assignment completeness of each tracer (sections~4.3--4.5 of \citealt{DESISV2024}). The increase over DR2 is only 6\% for the BGS, 8\% for the LRGs and 28\% for the ELGs, since completing the survey mainly adds footprint area outside the LSST overlap rather than density within it. We have checked that the resulting densities agree bin by bin, to within 10\%, with the five-year forecast table (table~7) of \citet{DESISV2024} for the tracers present in that table. We do not include the proposed DESI-II low-redshift program \citep{Schlegel2022}: its high-density sample covers only a ${\sim}1400\,\mathrm{deg}^2$ patch, which cannot be represented by a single homogeneous number density in the covariance.

For \textit{Euclid} we adopt the H$\alpha$ sample of the official forecast \citep[table~3 of][]{Blanchard2020}: 1268 galaxies~deg$^{-2}$ over $0.9<z<1.8$ with per-bin biases $b=1.46$--$1.90$. These counts are published as totals in four $\Delta z = 0.2$ bins, which we redistribute onto our $\Delta z = 0.1$ grid following the tabulated $\mathrm{d}N/\mathrm{d}z$ of their underlying model (model~3 from table~3 of \citealt[][]{Pozzetti2016}). 

For the galaxy bias we adopt DESI measurements, which apply equally to the DESI-like 4MOST selections: $b_{\mathrm{BG}}=1.34$ \citep{DESI2016} and $b_{\mathrm{LRG}}=1.93$--$2.31$ per redshift bin \citep{Yuan2024}, as also adopted by the 4MOST-CRS forecasts \citep{Verdier2025}, and $b_{\mathrm{ELG}}=1.3$.
Table~\ref{tab:lens_sample} lists the resulting per-bin effective number densities and $n$-weighted biases, which are the inputs to our covariance. The fiducial tomography comprises 16 top-hat lens bins: the range $0.1<z<0.4$ is split into four uniform sub-bins of width $\Delta z = 0.075$, followed by uniform $\Delta z = 0.1$ bins over $0.4<z<1.6$. The finer low-redshift binning keeps the closest lens--source pairs within the validity of the geometry-only approximation: with it, the iterative $\chi^2$ selection of Sec.~\ref{sec:sim_ratios_cov} discards no spectroscopic-sample combinations, at no cost in total signal-to-noise.
Fig.~\ref{fig:lens_sample} shows the contribution of each survey: 4MOST and DESI dominate at $z<0.9$, while \textit{Euclid} supplies most  lenses at $z>0.9$. 

We also note that lens magnification contamination would be harder to minimize in such a sample, since   spectroscopic selections do not in general sit near $\alpha_l = 1$, so this sample would require a different lens magnification characterization and mitigation than the photometric lens sample.

\begin{table*}
\centering
\caption{\textbf{Impact of systematic uncertainties in flat $w_0w_a$CDM.} 
$\mathrm{FoM}(w_0, w_a)$ for the lensing ratios (LR) alone and in combination with the CMB or  CMB + BAO + SNe (labelled as \textit{all}). Each row is labeled by the nuisance parameters marginalized over: ``photo-$z$'' samples one mean-redshift shift per tomographic bin with SRD-based Gaussian priors (17 lens + 9 source bins for LSST Y10; 9 source bins for the spectroscopic lens sample, whose lens redshifts are exact), and ``shear calib.'' one multiplicative bias per source bin. ``IA in data'' rows add an IA signal to the LR data vector (DES Y6 3x2pt best-fit TATT input) while keeping the IA-free geometric model; re-applying the $\chi^2 \leq 1$ selection to the contaminated data then retains a different set of ratios. The last row of each block combines all three effects. The values in parentheses give the gain from the ratios: for CMB+LR the FoM factor over the CMB alone ($\mathrm{FoM} = 2.92$), and for all probes the fractional improvement over CMB + BAO + SNe ($\mathrm{FoM} = 189.25$). FoM differences of ${\lesssim}2\%$ are within the sampling noise of the chains.}
\label{tab:lr_systematics}
\begin{tabular}{lrrr}
\hline
systematics & \multicolumn{1}{c}{LR} & \multicolumn{1}{c}{CMB+LR (gain)} & \multicolumn{1}{c}{all (gain)} \\
\hline
\multicolumn{4}{l}{\textit{LSST Y10 photo-$z$ lenses}} \\
    fixed & 8.7 & 59.0 (20.2$\times$) & 248 (+31\%) \\
    photo-$z$ & 5.6 & 31.2 (10.7$\times$) & 222 (+17\%) \\
    photo-$z$ + shear calib. & 3.9 & 23.7 (8.1$\times$) & 217 (+15\%) \\
    IA in data & 11.0 & 77.5 (26.5$\times$) & 253 (+34\%) \\
    IA in data + photo-$z$ + shear calib. & 4.1 & 23.6 (8.1$\times$) & 217 (+15\%) \\
\hline
\multicolumn{4}{l}{\textit{Spectroscopic lenses}} \\
    fixed & 5.0 & 23.5 (8.0$\times$) & 210 (+11\%) \\
    source photo-$z$ & 4.5 & 20.8 (7.1$\times$) & 209 (+10\%) \\
    source photo-$z$ + shear calib. & 3.7 & 18.6 (6.4$\times$) & 213 (+12\%) \\
    IA in data & 5.1 & 20.7 (7.1$\times$) & 208 (+10\%) \\
    IA in data + source photo-$z$ + shear calib. & 4.1 & 17.4 (5.9$\times$) & 205 (+8\%) \\
\hline
\end{tabular}
\end{table*}

Repeating the lensing ratios only analysis with this spectroscopic sample in place of the LSST Y10 lens sample yields $58\%$ of the FoM in the Flat $w_0w_a$CDM model, with fixed systematics (see Table~\ref{tab:lr_systematics}). The ratios' signal-to-noise budget is dominated by the CMB-lensing reconstruction noise and the source shape noise rather than by lens shot noise (Appendix~\ref{sec:error_budget}), so the $\sim 20\times$ drop in lens density costs far less proportionally. However, even if we find the cost  significant at fixed systematics,  the comparison changes once photo-$z$ uncertainty is included. Since the spectroscopic sample does not need lens photo-$z$ nuisance parameters, the gap narrows in this case, and we find the FoM for the lensing ratios alone is the same for both samples in the flat $w_0w_a$CDM model. When combined with the other geometrical probes, the photometric sample still performs slightly better, improving the CMB+BAO+SNe FoM by $15\%$ compared to $8\%$ for the spectroscopic sample. 

SPHEREx \citep{Dore2014}, an all-sky survey that began in 2025, could provide a further lens sample, as already considered for lensing ratios by \citet{BermejoCliment2020}. Its forecast galaxy samples are defined by disjoint ranges of redshift precision\footnote{Number densities and biases from \url{https://github.com/SPHEREx/Public-products}.}. The two most precise have redshift errors below our lens-bin widths ($\Delta z = 0.075$--$0.1$) and together reach ${\approx}0.96$~arcmin$^{-2}$ over a much larger area than the combined sample above. At $0.1<z<0.4$ this is more than three times the density of our combined sample, but $77\%$ of these galaxies lie at $z<0.6$, and at $1<z<1.6$ they provide ${\approx}0.04$~arcmin$^{-2}$, against $0.28$~arcmin$^{-2}$ in our combined sample. A SPHEREx sample would therefore complement the high-redshift \textit{Euclid} lenses rather than replace them. We leave a forecast with this sample to future work; like the other spectroscopic selections, it would need its own magnification characterization.

\section{Lensing-ratio error budget}\label{sec:error_budget}

\begin{figure*}
    \centering
    \begin{subfigure}[t]{0.49\textwidth}
        \centering
        \includegraphics[width=\textwidth]{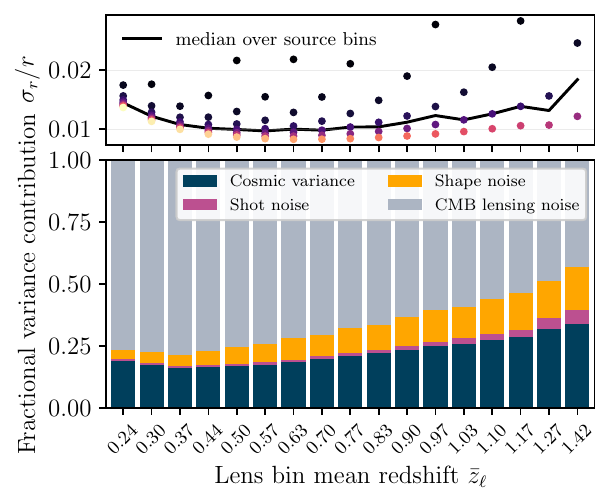}
        \caption{LSST Y10}
        \label{fig:error_budget_y10}
    \end{subfigure}
    \hfill
    \begin{subfigure}[t]{0.49\textwidth}
        \centering
        \includegraphics[width=\textwidth]{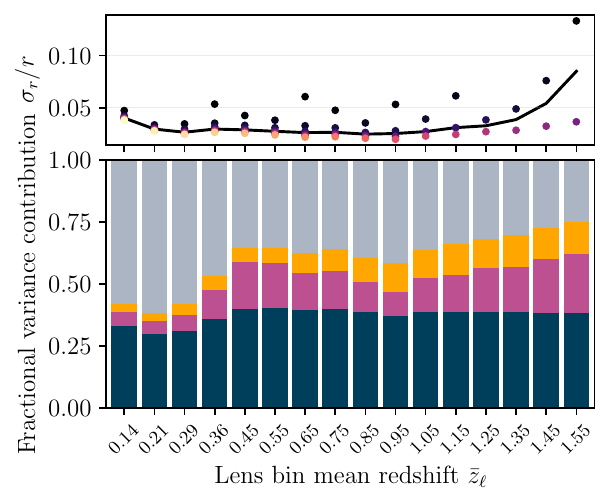}
        \caption{Spectroscopic lens sample}
        \label{fig:error_budget_specz}
    \end{subfigure}
    \caption{\textbf{Lensing-ratio error budget.} Noise budget of the lensing-ratio diagonal covariance for (a) LSST~Y10 and (b) the spectroscopic lens sample of App \ref{app:lens_specz}. Top: per-ratio fractional precision $\sigma_r/|r|$ for every lens-source combination, colour-coded by the lens--source separation $\bar z_s - \bar z_\ell$ (lighter is larger separation); the black line is the median over source bins. Bottom: fractional contribution of each of the four Gaussian noise sources to the total variance, combining source bins within each lens bin with an inverse-variance ($S/N^2$) weight so the budget reflects the ratios that actually carry the constraining power.}
    \label{fig:ratio_error_budget}
\end{figure*}

Figure~\ref{fig:ratio_error_budget} decomposes the lensing ratios covariance into its four Gaussian contributions: cosmic variance, galaxy shot noise, shear shape noise, and CMB-lensing reconstruction noise (Sec.~\ref{sec:covariance}). We plot this for every geometrically valid ratio, i.e.\ before the $\chi^2$ cut is applied. Recomputing the budget for the combinations retained by the $\chi^2 \leq 1$ cut shifts the per-lens-bin fractions by at most a few percentage points, so the figure describes the analysis set equally well. CMB-lensing reconstruction noise dominates the budget at low lens redshift, where the SO noise curve is smallest relative to the signal, and shot and shape noise grow in relative importance toward higher lens redshift as the galaxy samples become sparser.

\section{Impact of systematic uncertainties}\label{app:systematics}

We now switch on the leading systematic uncertainties one at a time and repeat the flat $w_0w_a$CDM analysis for both lens samples, for the ratios alone and in combination; Table~\ref{tab:lr_systematics} collects the resulting figures of merit. We adopt flat $w_0w_a$CDM as the reference model for this test because it is the space where the ratio-only constraints are strongest ($\mathrm{FoM}(w_0, w_a) = 8.7$, versus $\approx 1$ in the open models): the cost of each systematic is then more easily resolved in the ratio-only figures of merit, whereas in a plane that the ratios alone barely constrain the different treatments would be indistinguishable. The free-neutrino-mass model, where the ratios contribute most with respect to the current BAO+CMB+SNe baseline, is also revisited under the full marginalization at the end of this appendix. We consider first the photometric-redshift uncertainty of the lens and source samples, then also with multiplicative shear calibration, then only unmodelled intrinsic alignments, and finally all three effects together.

\paragraph{Photometric redshifts} The predictions are built from distance factors weighted by the tomographic $n(z)$s, see Eq.~\eqref{eq:ratio_cmb}, so a coherent shift of a bin's mean redshift changes them directly. We model this uncertainty with one additive mean-redshift shift per tomographic bin, $\Delta z_b$ (17 lens + 9 source parameters for LSST Y10), applied to the input $n(z)$ and marginalized over with zero-centred Gaussian priors. The prior widths start from the LSST Y10 forecast choices of \citet{Zhang2022}, $\sigma(\Delta z_b) = 0.003\,(1+\bar z)$ for the SRD lens sample and $0.001\,(1+\bar z)$ for the SRD sources, rescaled to our samples under the assumption that the calibration uncertainty scales 
inversely with the square root of the number of calibrating galaxies, $\sigma(\Delta z_b) \propto \sigma_z/\sqrt{N}$. For our bright-conservative lens sample, this scaling translates into two corrections, one per factor: an improvement by $0.02/0.03$, assuming the brighter selection improves the assumed photo-$z$ scatter from $0.03\,(1+z)$ to $0.02\,(1+z)$, and an inflation by $\sqrt{n_{\rm ref}/n_{\rm bin}}$, since our per-bin densities ($0.34$--$0.81$~arcmin$^{-2}$) are well below the SRD lens densities over the same redshift ranges ($2.6$--$4.3$~arcmin$^{-2}$). For the sources only the density correction applies: the same $27$~arcmin$^{-2}$ are split into 9 bins instead of 5, widening the priors by $\sqrt{5.4/3.0} = 1.34$. The net effect yields $\sigma(\Delta z_b) = 0.0059$--$0.0133$ for the lens bins and $0.0017$--$0.0045$ for the source bins (Table~\ref{tab:cases_updated}), wider than the  \citet{Zhang2022} values everywhere.  For LSST Y1 (13 lens + 5 source bins, Sec.~\ref{sec:results_y1}) we adopt the SRD Y1 requirements of \citet{Zhang2022} directly, $\sigma(\Delta z_b) = 0.005\,(1+\bar z)$ for the lenses and $0.002\,(1+\bar z)$ for the sources ($0.0058$--$0.0097$ and $0.0025$--$0.0051$), together with $\sigma_m = 0.013$ per source bin. No rescaling is applied, since the Y1 sources are the standard 5-bin SRD split.

Marginalizing over these  shifts, the ratio-only $\mathrm{FoM}(w_0, w_a)$ goes from $8.7$ to $5.6$, and the CMB+LR combination from $59$ to $31$ (Table~\ref{tab:lr_systematics}). Once BAO and SNe are added it goes from $248$ to $222$. Figure~\ref{fig:pz_posteriors} compares the shift posteriors with their priors: the ratios self-calibrate the lens shifts, constraining them to $0.36$--$0.79$ of the prior width, most strongly at high redshift where the prior is widest, while the source shifts remain prior-dominated ($0.7$--$1.0$ of the prior width).  The spectroscopic lens sample does not need the lens redshift marginalization; marginalizing over the sources only costs the spectroscopic ratios just $10\%$ ($5.0$ to $4.5$), $12\%$ in the CMB+LR combination ($23.5$ to $20.8$), and leaves the full combination unchanged within the sampling noise. Under realistic photo-$z$ uncertainty the FoM gap between the two lens samples therefore narrows but the photometric sample still wins.

\paragraph{Shear calibration} The multiplicative shear-calibration parameters $m_j$ of Eq.~\eqref{eq:ratio_model_sys}, one per source bin, are treated the same way: zero-centred Gaussian priors of width $\sigma_m = 0.003 \times \sqrt{5.4/3.0} = 0.0040$, the SRD requirement adopted by \citet{Zhang2022} inflated by the same calibration-statistics factor for our 9-bin split of their 5-bin source configuration. Adding the shear-calibration marginalization on top of the photo-$z$ shifts has a significant effect: the LSST Y10 ratio-only FoM drops from $5.6$ to $3.9$, the CMB+LR combination from $31.2$ to $23.7$, while the change in the full combination, ${\sim}2\%$, is at the level of the sampling noise. The spectroscopic sample behaves the same way ($4.5$ to $3.7$, $20.8$ to $18.6$, full combination unchanged within sampling noise). Then, the gap between the two lens samples narrows further though the photometric sample still remains more competitive. 

\paragraph{Intrinsic alignments} Intrinsic alignments enter the ratios only through the galaxy--shear correlations: the alignment contribution adds to $\gamma_t$ where the source redshift distribution overlaps the lens, while the galaxy--CMB-lensing correlations are unaffected. To quantify the impact we generate data vectors with a nonzero input alignment signal, the TATT model \citep{Blazek2019} at the DES Y6 3x2pt \citep{DESY6Cosmo2026} best-fit parameters (with tidal-alignment and tidal-torquing amplitudes $A_1 = 0.539$ and $A_2 = -0.889$, redshift-evolution indices $\eta_1 = -0.363$ and $\eta_2 = -1.040$), and analyze them with the alignment-free geometric model. The $\chi^2 \leq 1$ selection of Sec.~\ref{sec:sim_ratios_cov}, applied to the contaminated data, removes the strongly contaminated lens--source-adjacent combinations and reshuffles the kept set relative to the fiducial one: for LSST Y10 it newly discards eight high-redshift combinations while recovering nine low-redshift ones that the fiducial cut removes (60 of 84 kept; crosses in Fig.~\ref{fig:datavector_basis1}), and for the spectroscopic sample it discards fifteen (64 of 79 kept).  LSST Y10's FoM  improves slightly (ratio-only $11.0$ versus $8.7$, CMB+LR $77.5$ versus $59.0$; Table~\ref{tab:lr_systematics}) because the recovered low-redshift combinations carry more information than the discarded ones. We also verified that recomputing the covariance with the intrinsic alignment contributions included changes the errors of the retained ratios by less than $0.3\%$  and leaves the selection unchanged, so the fiducial covariance is used throughout.

\paragraph{Combined marginalization} The last row of each block is the realistic bottom line of this forecast: IA-contaminated data analyzed with the photo-$z$ and shear-calibration uncertainties marginalized simultaneously. Relative to the ideal fixed-systematics configuration, the ratios alone retain about half of their constraining power for LSST Y10 ($8.7 \to 4.1$) but four fifths for the spectroscopic sample ($5.0 \to 4.1$); in both cases the cost is driven by the calibration priors rather than by the intrinsic alignments contamination. Under this full treatment the headline conclusions are preserved: the ratios still improve the CMB-alone figure of merit by $8.1\times$ ($5.9\times$ spectroscopic) and the full CMB + BAO + SNe combination by $+15\%$ ($+8\%$).

\paragraph{Other models} Here we check whether the conclusions obtained for the flat $w_0w_a$ model generalize to other models. In particular we check  the $w_0w_a$CDM + $\Omega_k$+$\sum m_\nu$ model, where the lensing ratios gains are the largest and thus potentially one of the most interesting cases. We find that adding the ratios to CMB + BAO + SNe there yields $\mathrm{FoM}(w_0, w_a) = 134$ versus $161$ at fixed systematics, still a $+21\%$ improvement over the LR-free baseline of $111$, so that is a similar reduction of the flat model. Adding the IA contamination on top of the same marginalization, so the final configuration, with all three systematic effects active, leaves this essentially unchanged: $\mathrm{FoM}(w_0, w_a) = 132$, a $+19\%$ gain over the baseline.

\paragraph{CMB-lensing systematics} The systematics considered so far enter through the galaxy-lensing measurements; the reconstructed CMB convergence map carries its own. The most important are extragalactic foregrounds in the CMB maps: the thermal Sunyaev--Zel'dovich (tSZ) effect, the cosmic infrared background, and unresolved point sources. These fields trace the same structure as the lens galaxies, so they do not average down in the cross-correlation: they add a spurious contribution to $\xi_{g\kappa_{\rm CMB}}$, and because $\kappa_{\rm CMB}$ enters only the numerators of the CMB ratios, the bias propagates directly to the ratios instead of cancelling. The tSZ contamination, which is largest at small scales, was a limiting systematic of the measurement in \citetalias{PratBaxter2019}~\citep{PratBaxter2019}. However their mitigation is by now standard on the map side: foreground-deprojected and bias-hardened reconstruction largely removes the tSZ contribution at a modest noise cost  \citep{MadhavacherilHill2018, Sailer2020, Darwish2021}, and polarization-based reconstruction, which becomes competitive at SO sensitivity, is essentially immune to these foregrounds. We assume residual CMB-lensing foreground biases can be
controlled to a negligible level. The adopted noise curves
do not include a penalty for foreground mitigation.

\section{Impact of the source-sample optimization}\label{app:source_opt}\label{sec:results_source_opt}

\begin{figure}
    \centering
    \includegraphics[width=\columnwidth]{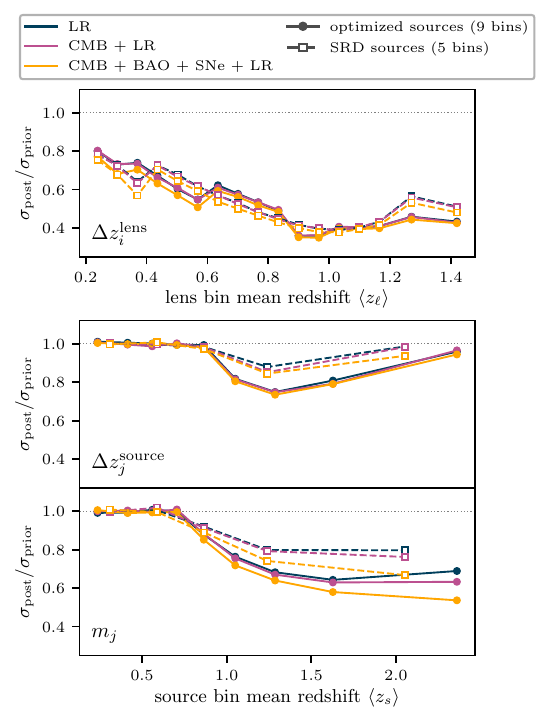}
    \caption{\textbf{Self-calibration of the nuisance parameters.} Ratio of the posterior to the prior standard deviation under the full marginalization (photo-$z$ shifts and shear calibration simultaneously), in flat $w_0w_a$CDM, for the LSST Y10 lens mean-redshift shifts (top), source shifts (middle) and multiplicative shear biases (bottom), as a function of the mean redshift of each bin. Filled circles with solid lines show the optimized 9-bin source sample, and open squares with dashed lines the standard SRD 5-bin sample, both paired with the same 17 lens bins; colors indicate the data combination, and the dotted line marks a prior-dominated parameter. The ratios self-calibrate the lens shifts to $0.35$--$0.80$ of the prior width for both source samples. The source shifts and shear biases are prior-dominated for $\langle z_s\rangle < 0.9$ and self-calibrate at higher redshift, more strongly for the optimized split. }
    \label{fig:pz_posteriors}
\end{figure}

\begin{table*}
\centering
\caption{\textbf{Impact of the source-sample optimization.} Flat $w_0w_a$CDM $\mathrm{FoM}(w_0, w_a)$ for the optimized 9-bin source split used throughout this work versus the standard SRD Y10 source sample (5 equal-number bins of the same $27$~arcmin$^{-2}$). Each row is labeled by the nuisance parameters marginalized over (``fixed'' = none); ``IA in data'' rows fit the IA-contaminated data vectors with the unchanged IA-free model, after each configuration's own $\chi^2 \leq 1$ selection. For the ratios alone and for all probes combined (CMB+LR+BAO+SNe) we give the FoM for the optimized and the standard source sample; for all probes, the percentage in parentheses is the fractional improvement over the CMB+BAO+SNe baseline ($\mathrm{FoM} = 189.25$).}
\label{tab:source_opt}
\small
\setlength{\tabcolsep}{4pt}
\begin{tabular}{lrrrr}
\hline
 & \multicolumn{2}{c}{LR} & \multicolumn{2}{c}{all (gain)} \\
systematics & \multicolumn{1}{c}{optimized} & \multicolumn{1}{c}{SRD} & \multicolumn{1}{c}{optimized} & \multicolumn{1}{c}{SRD} \\
\hline
\multicolumn{5}{l}{\textit{LSST Y10 photo-$z$ lenses}} \\
    fixed & 8.7 & 8.3 & 248.4 (+31\%) & 251.8 (+33\%) \\
    photo-$z$ + shear calib. & 3.9 & 3.8 & 217.0 (+15\%) & 216.0 (+14\%) \\
    IA in data + photo-$z$ + shear calib. & 4.1 & 3.9 & 217.0 (+15\%) & 217.0 (+15\%) \\
\hline
\multicolumn{5}{l}{\textit{Spectroscopic lenses}} \\
    fixed & 5.0 & 4.2 & 210.2 (+11\%) & 206.9 (+9\%) \\
    source photo-$z$ + shear calib. & 3.7 & 3.4 & 212.7 (+12\%) & 211.7 (+12\%) \\
    IA in data + source photo-$z$ + shear calib. & 4.1 & 3.0 & 204.8 (+8\%) & 204.9 (+8\%) \\
\hline
\end{tabular}
\end{table*}

The narrow 9-bin source sample adopted throughout this work was designed to maximize the constraining power of the lensing ratios. To quantify what this optimization buys, we repeat the analysis with the standard SRD Y10 source sample instead: the same $27$~arcmin$^{-2}$ split into 5 equal-number bins (dashed outlines in Fig.~\ref{fig:nzs}). Note that only the source sample is swapped in this test: the lenses remain the optimized samples adopted in Sec.~\ref{sec:data}, the modified 17-bin photometric selection and the spectroscopic sample, rather than the standard SRD lens definition. 

At the signal-to-noise level the two source configurations are nearly identical: $368$ for the standard versus $360$ for optimized for the photometric lens sample, and $167$ versus $166$ for the spectroscopic one. These S/N are computed  after each configuration's own $\chi^2 \leq 1$ geometry-validity selections. 

However, equal signal-to-noise does not imply equal parameter leverage. The SRD tomographic bins also carry fewer and better-calibrated nuisance parameters. Moreover, the surviving ratios sample the lens--source geometry differently in the two configurations, so the actual impact of the optimization must be assessed on the cosmology posteriors. Table~\ref{tab:source_opt} collects the ratio-only figures of merit across the systematics ladder of Appendix~\ref{app:systematics}. At fixed systematics the optimization buys already not much for the photometric lens sample ($8.7$ versus $8.3$) and similarly for the spectroscopic one ($5.0$ versus $4.2$). Then, when marginalizing over photo-$z$ and shear systematics, the standard sources yield even closer FoM to the optimized sample.

The one regime where the optimization still helps appreciably is the IA-contaminated ratios with spectroscopic lenses ($3.0$ versus $4.1$), where the coarser 5-bin $\chi^2 \leq 1$ selection discards more of the usable signal. Still, also this difference washes out once external data are added: with all probes combined the two samples are indistinguishable ($204.9$ versus $204.8$). The optimized sample therefore only matters for ratios-alone forecasts; under realistic marginalization, or in combination with CMB and other background probes, the standard SRD source sample performs essentially as well. Thus, we conclude that there will be no need to generate an optimized \textit{source} sample for the lensing ratios probe and we can use the standard SRD sources.

We also check that these conclusions are not specific to flat $w_0w_a$CDM: repeating the photometric-lens comparison in open $w_0w_a$CDM, the standard sources retain $98\%$ of the ratios-only FoM ($1.14$ versus $1.16$) and $94\%$ with the compressed CMB prior ($10.2$ versus $10.9$); in open $\Lambda$CDM the two samples are indistinguishable; and the free-neutrino-mass result is unchanged both at fixed systematics ($161.3$ versus $161.4$, preserving the $+45\%$ gain exactly) and under the full photo-$z$ + shear-calibration marginalization ($133.6$ versus $134.0$, a $+20\%$ versus $+21\%$ gain over the baseline).

The same conclusion holds for the non-parametric reconstruction of Sec.~\ref{sec:results_binned_de}. To minimize computational costs for this side test we make this comparison with a Fisher approximation: the geometric ratio model of Sec.~\ref{sec:ratios} linearized around the fiducial, with the external information taken from the sampled baseline posteriors. Validated against the corresponding optimized-sample chains, the forecast reproduces the per-bin uncertainties to better than $11\%$, always slightly underestimating them. We find that for the $\theta_\star$ + BAO + SNe combination with free curvature, where the ratios matter most, the two source samples are nearly equivalent: a mean improvement of $39.5\%$ versus $42.4\%$ with systematics fixed, $31.2\%$ versus $33.4\%$ marginalized, and $\sigma(\Omega_k) = 0.0054$ versus $0.0052$. The difference is concentrated in the last redshift bin ($46\%$ versus $54\%$, when marginalizing over systematics), where the coarser source binning forces the intrinsic alignment selection to discard more of the high-redshift signal.

Figure~\ref{fig:pz_posteriors}  compares the nuisance-parameter posteriors of the two source samples under the full marginalization.   Both samples self-calibrate the 17 (shared) lens shifts in the same way, while the source shifts and shear biases remain close to their priors, with mild self-calibration at the highest source redshifts that is somewhat stronger for the optimized sample.

\bibliography{library.bib}

\end{document}